\documentclass[apl,twocolumn,floatfix,amsfonts,showpacs]{revtex4}
\usepackage{graphicx,graphics,color,epsfig}
\usepackage{bm}
\usepackage{amsmath}
\usepackage{amssymb}

\begin{document}

\title{Conductance fluctuation and shot noise in disordered graphene systems, a perturbation expansion approach}

\date{\today}

\author{Jia Ning Zhuang and Jian Wang$^*$}

\affiliation{Department of Physics and the center of theoretical and
computational physics,\\The University of Hong Kong, Pokfulam Road,
Hong Kong, China}

\begin{abstract}
We report the investigation of conductance fluctuation and shot noise in disordered graphene systems with two kinds of disorder, Anderson type impurities and random dopants. To avoid the brute-force calculation which is time consuming and impractical at low doping concentration, we develop an expansion method based on the coherent potential approximation (CPA) to calculate the average of four Green's functions and the results are obtained by truncating the expansion up to 6th order in terms of ``single-site-T-matrix". Since our expansion is with respect to ``single-site-T-matrix" instead of disorder strength $W$, good result can be obtained at 6th order for finite $W$.
We benchmark our results against brute-force method on disordered graphene systems as well as the two dimensional square lattice model systems for both Anderson disorder and the random doping. The results show that in the regime where the disorder strength $W$ is small or the doping concentration is low, our results agree well with the results obtained from the brute-force method. Specifically, for the graphene system with Anderson impurities, our results for conductance fluctuation show good agreement for $W$ up to $0.4t$, where $t$ is the hopping energy. While for average shot noise, the results are good for $W$ up to $0.2t$. When the graphene system is doped with low concentration $1\%$, the conductance fluctuation and shot noise agrees with brute-force results for large $W$ which is comparable to the hopping energy $t$. At large doping concentration $10\%$, good agreement can be reached for conductance fluctuation and shot noise for $W$ up to $0.4t$. We have also tested our formalism on square lattice with similar results. Our formalism can be easily combined with linear muffin-tin orbital first-principles transport calculations for light doping nano-scaled systems, making prediction on variability of nano-devices.
\end{abstract}

\maketitle

\section{Introduction}
In nano-electronics, quantitative evaluation of impurity effects is crucial because device properties are strongly influenced by or even built on such effects. Experimentally, the impurities exist and can be doped in nano-devices without knowing their exact locations, so theoretically it is important to predict the averaged transport quantities such as conductance over impurity configurations. The most direct way to obtain the averaged conductance is to generate many different configurations, then calculate the conductance for each configuration, and finally take the mean value. Such a brute-force method is usually used in the mesoscopic systems from diffusive regime to localized regime because it is an exact calculation. But in order to get good statistics, huge number of configurations has to be generated making it very time-consuming especially for the calculation of conductance fluctuation. When the disorder strength is weak, it is not necessary to use the brute-force method since some analytic approximate method is superior in speed while maintaining the same accuracy. For this purpose considerable effort has been made to develop approximate techniques, within which the most widely used technique is the coherent potential approximation (CPA), which is a useful tool to evaluate the configurational averaged one-electron Green's function\cite{soven_cpa} $\langle G\rangle$,
and has also been extended to determine the so-called ``vertex corrections"\cite{velicky_cpa} for quantities
involving two Green's functions. CPA approach has been implemented in the
Korringa-Kohn-Rostoker\cite{kkr1,kkr2,kkr3} and linear muffin-tin orbital\cite{lmto1,lmto2,lmto3} for first principles calculations and has many successful applications\cite{cpa_appl1,cpa_appl2,cpa_appl3}.
The central idea of CPA is to find a ``coherent potential" such that the one-electron Green's function evaluated under such potential approximately equals the configurational averaged Green's function. As an extension, CPA can also be used to determine the so-called ``vertex corrections"\cite{velicky_cpa} for the product of two Green's functions. Later,
Levin \textit{et al} also proposed an elegant diagrammatic method to evaluate the Hall coefficient which relates to the direct multiple of three Green's functions\cite{velicky_diagram}.
Importantly, the CPA approach and its extensions can be combined with local-orbital based DFT to calculate the physical properties, such as the band structure and the density of states, of realistic materials. One example is the development of the so called ``KKR-CPA", used to study the band structure and density of states of
Cu-Ni\cite{kkr1}, Ag-Pd\cite{kkr2}, and Cu-Pd\cite{kkr3} alloys. The linear muffin-tin orbital (LMTO) method has also
been proposed\cite{lmto1} and used to study the electronic structures of metal alloys\cite{lmto2,lmto3}. CPA combined with LMTO works very well and has many successful applications. Examples are the investigation of transport properties in disordered magnetic multilayers\cite{cpa_appl1}, structure of Sn-Ge alloys\cite{cpa_appl2}, the electronic structure of non-stoichiometric compounds\cite{cpa_appl3}, and doped semiconductors\cite{wy_apl,mathiew_apl}.

The latest development of CPA extended its range of application to non-equilibrium quantum transport problems where impurity average has to be performed. One prominent work is the ``non-equilibrium vertex correction" (NVC) discussed in Ref.[\onlinecite{ke_cpanvc}]. It has been shown by Zhuravlev \textit{et al.}\cite{cpa_voltage_probe} that this NVC formalism can be interpreted in terms of the B\"uttiker voltage-probe model so that it is not merely a correction to
the electronic structure.\cite{velev_2012} Generally speaking, this site-oriented algorithm to evaluate the
average conductance is well developed and adopted by different groups.\cite{NVC_app1,NVC_app2}

In the presence of disorder CPA-NVC approach allows one to calculate non-equilibrium transport properties
such as I-V curve and other quantities involving two Green's functions. However, it can not be applied directly
to investigate equilibrium transport properties involving four Green's functions such as conductance
fluctuation and shot noise. Since the fluctuation of transport properties of nano-devices, known as "variation" of nano-devices, is a very important quantity in nano-electronics and it provides the information on how much the specific device configuration could deviate from the mean value. We notice that a quantified experiment has been reported to measure such kind of fluctuation\cite{FLYang_fluct_exp} recently. Therefore, it is timely to develop a theoretical formalism that is capable of treating disorder average of four Green's functions.
To the best of our knowledge, so far this is still an outstanding problem yet to solve based on CPA approach. One possible reason is that, the NVC could be regarded as a perturbation expansion approach based on CPA to evaluate the conductance by including the ladder diagrams. For conductance fluctuation, however, such a partial summation is not good enough. In this paper, we develop a direct perturbation expansion with respect to the ``single-site-T-matrix" up to a given order which is a good approximation for weak disorder strength or small doping concentration. We carry out benchmark calculation of average conductance, shot noise, and conductance fluctuation using the direct expansion method on a graphene system and a two-dimensional lattice model with Anderson impurities as well as random dopants. We have  compared our results with the brute-force calculation. We find that a six-order expansion can give very good results for conductance fluctuation and shot noise when disorder strength $W$ is comparable to the hopping strength $t$, $W \sim 0.4t$. In the presence of doping, our results also show good agreement with that obtained from brute-force method at low doping concentration. We note that our method can be easily implemented in the first principles transport calculation in nanostructures.


The rest of this paper is organized as the following. In section II, we briefly revisit CPA formalism and introduce our direct expansion approach to calculate disorder average of four Green's functions. An expansion view on NVC method is also provided. In section III, we compare our results with that obtained from the brute-force method on a graphene system and square lattice of size $40 \times 40$ for two types of disorder: Anderson disorder and different doping concentration. The results for average conductance, shot noise\cite{buttiker_shotnoise}, and the conductance fluctuation are also presented. Finally we conclude our work in section \ref{conclusion}.

\section{Theoretical formalism}
\label{formalism}

We consider a tight-binding mode on a square lattice model described by the following Hamiltonian:
\begin{eqnarray}\label{H_cent}
H_{c} = \sum_i (4t+v_i) c_i^\dagger c_i - t \sum_{<ij>} c_i^\dagger c_j
\end{eqnarray}
where $t$ is the nearest neighbor hopping energy and $c_i$ and $c_i^\dagger$ are electron annihilation and creation operators on atomic site $i$ respectively. We choose $t=1$ as the energy unit. The on-site energy chosen as $4t$ is a convention that the energy bottom of the 2D band structure to be zero.

We also assume that the structure of the left and right leads has a similar interaction. The effect of leads can be taken into account by self-energy\cite{datta_book} $\Sigma_L^{r,a}$ for the left and $\Sigma_R^{r,a}$ for the right. The self-energy of leads can be calculated numerically\cite{selfenergy1,selfenergy2}. Although the Hamiltonian in Eq.(\ref{H_cent}) is very simple, our direct expansion in principle can handle more complicated Hamiltonians as long as it only contains single particle interactions. It is also straightforward to generalize our approach to the case of multi-orbital per site.
Here we consider ``diagonal disorder"\cite{shengping_book} with  disorder strength $v_i$ on $i$th atomic site. Different types of disorder can be described by introducing a ``probability function" for $v_i$. We consider two different types of disorder. One is ``Anderson disorder" with the probability function given by
\begin{equation}\label{rho_anderson}
\rho(v_i) = \begin{cases}
1/w, & -w/2 \leq v_i \leq w/2,  \\
0, & \text{otherwise}
\end{cases} \qquad \forall i \quad \mathrm{in\quad center},
\end{equation}
where $w>0$ is called the strength of Anderson disorder. Another is to dope the system with different type of atom:
\begin{equation}\label{rho_dope}
\rho(v_i) = p \delta(v_i-w) + (1-p) \delta(v_i-0).
\end{equation}
Here $0\leq p\leq 1$ is the doping concentration, and $w$ is the energy difference between the dopant and the original atom. In the theoretical formalism we can general types of diagonal disorder including these two types of disorder.

\subsection{CPA Algorithm}
In this subsection, we revisit the well-developed ``single-site CPA", because this is the starting point of our direct  expansion approach.
CPA is an approximation to evaluate the averaged single-particle retarded or advanced Green's function ($\langle G^{r}\rangle$ or $\langle G^{a}\rangle$), and it is known to be good in homogeneous ensembles\cite{shengping_book}. In realistic nano-devices with small concentration, it has been shown that the NVC which based on CPA also works very well.\cite{ke_cpanvc}.
In CPA approximation, the disorder effect renormalizes the on-site energy by adding a ``coherent potential" $(\hat{\Delta E}) = \sum_i (\Delta E)_{i}|i\rangle\langle i|$ on each atomic site, such that
\begin{eqnarray}\label{cpa_equation}
\langle G^{r}\rangle = G_e^r
\end{eqnarray}
where $G_e^r$ denotes equilibrium Green's function in the absence of disorder and can be expressed as
\begin{eqnarray}
G_e^r=[(E-\hat{\Delta E})-H_{c} - \Sigma^r(E)+i\eta]^{-1},
\end{eqnarray}
in which $\Sigma^r = \Sigma^r_L+\Sigma^r_R$ is the total self-energy due to the leads, and $\eta$ is a infinitesimal positive number.

For a given disorder configuration, the Green's function $G^r$ is related to $G_e^r$ by a ``T-matrix",
\begin{equation}\label{Tmat}
G^r=G_e^r+G_e^r  T^r G_e^r.
\end{equation}
in which the ``T-matrix" is used to describe one specific disorder configuration, and it can also be understood as the ``irreducible" self-energy induced by the disorder.
Taking configurational average on both sides, and compare with Eq.(\ref{cpa_equation}), we require \begin{equation}\label{general_cpa_condition}
\langle T^r \rangle = 0.
\end{equation}
However, to implement CPA, we need a further approximation, which is usually referred as ``weak overall scattering approximation" or ``single-site approximation", and either of them can lead to the CPA condition:
\begin{equation}\label{cpa_condition}
\langle T^r_i \rangle = 0,
\end{equation}
where $T^r_i$ is a matrix with only one non-vanishing element
\begin{equation}\label{Ti_component}
T^r_i=\tau^r_i|i\rangle\langle i|,
\end{equation}
and $\tau^r_i = \{[v_i-(\hat{\Delta E})_i]^{-1}-(G_e^r)_{ii}\}^{-1}$. Taking average on T-matrix, we have
\begin{equation}
\langle T^r_i\rangle = |i\rangle\langle i|\int  \frac{\rho(v) dv}{[v-\hat{\Delta E}_i]^{-1}-G_{e,ii}^r}=0
\end{equation}
from which the self-consistent equation for $(\hat{\Delta E})_{i}$ can be obtained \cite{shengping_book}
\begin{equation}
\hat{\Delta E_i} = \int  \frac{\rho(v) v dv}{[1-G_{e,ii}^{r}(v -\hat{\Delta E_i})]}.
\end{equation}
This equation is easy to converge.

\subsection{Direct Expansion}

With the definition of the linewidth function, $\Gamma_{L,R} = i(\Sigma_{L,R}^r-\Sigma_{L,R}^a) $, we can define the transmission matrix $\mathcal{T}=G^r\Gamma_L G^a\Gamma_R$. The averaged conductance (set $2e^2/h=1$) is
defined as $\langle\mathrm{Tr}(\mathcal{T})\rangle$, and the averaged DC shot noise is proportional to
$\langle\mathrm{Tr}(\mathcal{T}-\mathcal{T}^2)\rangle$, while the conductance fluctuation reads
$\sqrt{\langle[\mathrm{Tr}(\mathcal{T})]^2\rangle-\langle\mathrm{Tr}(\mathcal{T})\rangle^2}$. The averaged conductance
is usually calculated within the NVC approximation. While the shot noise and conductance fluctuation involve four Green's function and NVC approach can not apply here. Our direct expansion approach is to expand them according to Eq.(\ref{Tmat}), together with the T-matrix expansion with respect to single-site-T-matrix $T^r_i$ as the following:
\begin{eqnarray}\label{Tmat_expansion}
&&T^r = \sum_i T^r_i + \sum_{j\neq i}T^r_i G_e^r T^r_j+\sum_{j\neq i}\sum_{k\neq j}T^r_i G_e^r T^r_j G_e^r T^r_k \nonumber \\
&&+\sum_{j\neq i}\sum_{k\neq j}\sum_{l\neq k}T^r_i G_e^r T^r_j G_e^r T^r_k G_e^r T^r_l+\cdots
\end{eqnarray}
Notice that the multiple summation in Eq.(\ref{Tmat_expansion}) requires that the successive index should not be the same. Plugging this expansion into the expression of conductance fluctuation or shot noise and generate all the diagrams up to a certain order in $T_i$ and then store them once for all. Here we think $T_i$ is a natural expansion parameter because it describes the on-site scattering and it is a small quantity under small disorder strength and low doping concentration. With all the diagram generated, the average value of shot noise and conductance fluctuation can be calculated for different systems numerically.

\subsubsection{A. Averaged Shot Noise}
Considering the ${\cal T}^2$ term in the DC shot noise that involves the average of four Green's functions $\langle G^r \Gamma_L G^a \Gamma_R G^r \Gamma_L G^a\rangle$, we substitute Eq.(\ref{Tmat}) into this expression and it generates several terms up to the fourth order in T-matrix. The terms with only one T-matrix vanish due to Eq.(\ref{general_cpa_condition}). In the following we illustrate how to use direct expansion method to generate diagrams for the other terms involving multi-T-matrices.

As an example, one typical term containing three T-matrices is $G_e^r\langle T^r G_e^r \Gamma_L G_e^a T^a G_e^a \Gamma_R G_e^r T^r\rangle G_e^r \Gamma_L G_e^a$. We focus on the average part $\langle T^r X_1 T^a X_2 T^r\rangle$, where $X_1 = G_e^r \Gamma_L G_e^a$ and $X_2 = G_e^a \Gamma_R G_e^r$ are independent of randomness. We expand this average using Eq.(\ref{Tmat_expansion}) and truncate the resulting series to a certain order in $T_i$ (we have obtained 8th order). For this three T-matrices term the lowest order in $T_i$ is three because there is no zero-order term in Eq.(\ref{Tmat_expansion}), and all higher order terms (we will call them diagrams from now on) in $T_i$ up to our target order can be generated. {\it Symbolically}, we write
\begin{equation}
\mathrm{Tr}[\langle T^r X_1 T^a X_2 T^r X_3\rangle] = \sum_{n,m,l} C_{n,m,l} (T_i^r)^n (T_j^a)^m (T_k^r)^l
\end{equation}
This equation is symbolic so there is no summation over site indices i, j, and k. Here $C_{n,m,l}$ represents all the diagrams with the same order of $(n,m,l)$ in $T_\alpha$ ($\alpha=i,j,k$) contributed from different site indices i,j,k. Since $T_i$ is a matrix and does not commute with $X_{1/2/3}$, we have to keep both indices $n,l$. Obviously, we need to find two things: (1) how many combinations of $(n,m,l)$ we have; (2) how many diagrams are there for a particular $(n,m,l)$ due to different site indices $i,j,k$.

For instance, up to the 6th order ($n+m+l=6$), we have $(123)$, $(114)$ along with all their permutations and $(222)$, totally 10 different combinations. $C_{n,m,l}$ can be calculated by counting different combinations of $i,j,k$ and for $(n,m,l)=(132)$ it is obtained from the following expression,
\begin{eqnarray}\label{expand_2}
&&\sum_{i_1}\sum_{k_2\neq j_2, j_2\neq i_2}\sum_{j_3\neq i_3}\nonumber \\
&&\langle \mathrm{Tr}[T_{i_1}^r X_1 T_{i_2}^a G_e^a T_{j_2}^a G_e^a T_{k_2}^a X_2 T_{i_3}^r G_e^r T_{j_3}^r X_3]\rangle,
\end{eqnarray}
which is a six-multiple summation and can be handled using single-site CPA.

The evaluation of disorder average of Eq.(\ref{expand_2}) seems to be impossible. However, we note that each $T_i$ is a matrix with only one matrix element (it becomes a diagonal block matrix in the multi-orbital case, e.g., if spin-orbit interaction is considered), as in Eq.(\ref{Ti_component}). This simplifies calculation drastically. In addition, the CPA condition Eq.(\ref{cpa_condition}), indicates that if the summation index appears only once, the average vanishes.

So we have to find out all possible combinations of those six site indices, and there are many possibilities. For example, we can have $i_1=i_2,j_2=k_3,k_2=j_3$, and this combination gives the following contribution to Eq.(\ref{expand_2})
\begin{eqnarray}\label{T3_eg1}
&&\sum^{'}_{ijk}\langle \mathrm{Tr}[T_{i}^r X_1 T_{i}^a G_e^a T_{j}^a G_e^a T_{k}^a X_2 T_{k}^r G_e^r T_{j}^r X_3]\rangle \nonumber \\
&&=\sum'_{ijk}(X_1)_{ii} (G_e^a)_{ij} (G_e^a)_{jk} (X_2)_{kk} (G_e^r)_{kj} (X_3)_{ji} \nonumber \\
&& \times \langle\tau_i^r\tau_i^a\rangle\langle \tau_j^r\tau_j^a\rangle\langle\tau_k^r\tau_k^a\rangle,
\end{eqnarray}
where the prime on top of $\sum$ means the indices in the summation are mutually different and $\tau^r_i$ is defined after Eq.(\ref{Ti_component}). Another possible combination is $i_1=j_2=j_3,i_2=k_2=j_3$ which gives
\begin{eqnarray}\label{T3_eg2}
&&\sum'_{ij}\langle \mathrm{Tr} T_{i}^r X_1 T_{j}^a G_e^a T_{i}^a G_e^a T_{j}^a X_2 T_{j}^r G_e^r T_{i}^r X_3\rangle \nonumber \\
&&=\sum'_{ij}(X_1)_{ij} (G_e^a)_{ji} (G_e^a)_{ij} (X_2)_{jj} (G_e^r)_{ji} (X_3)_{ii}\nonumber \\
&& \times \langle(\tau_i^r)^2\tau_i^a\rangle\langle\tau_j^r(\tau_j^a)^2\rangle.
\end{eqnarray}

\begin{figure}[htbp]\vspace{0.2cm}
\includegraphics[height=1.8in,width=3in]{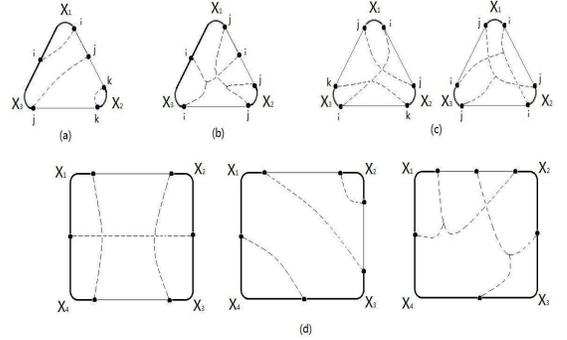}%
\caption{Typical diagrams included in the evaluation of shot noise. (a) The diagram corresponding to Eq.(\ref{T3_eg1}). (b) The diagram corresponding to Eq.(\ref{T3_eg2}). (c) Examples of other sixth order diagrams on $3T$ terms. (d) Examples of sixth order diagrams on $4T$ terms. }\label{diagram_shotnoise}
\end{figure}

Alternatively, we can have a much simpler diagrammatic representation of our expansion on the averaged shot noise. This representation is very similar to that of Levin \cite{velicky_diagram}. As an example, Eq.(\ref{T3_eg1}) can be diagrammatically expressed as Fig.\ref{diagram_shotnoise}(a) while the diagram corresponding to Eq.(\ref{T3_eg2}) is shown in Fig.\ref{diagram_shotnoise}(b). The thick lines in diagrams of Fig.\ref{diagram_shotnoise} represent the known matrix $X_i$, and the black dots represent the single site T-matrix $T_i$. Diagrammatically, expansion up to sixth order means that we only take into account those diagrams with the number of such black dots less than six. The thin line between two black dots represents either $G_e^r$ or $G_e^a$, depending on the configuration. The site indices such as $i$, $j$ and $k$ should be different one from another, and we should also keep in mind that the indices of two ends of a thin line can not be identical, from Eq.(\ref{Tmat_expansion}). Furthermore, we have to connect the repeated site
indices with the dashed lines, like Fig.\ref{diagram_shotnoise}(d) when we have four T matrices. By constructing such a diagrammatic rule, our expansion can be carried out by finding all the topologically distinct diagrams in which the number of black dots(single site T-matrix) is not more than six. Numerically, this procedure can be implemented by computer from which we can calculate the average conductance and shot noise.

\subsubsection{B. Conductance fluctuation}

Comparing with the averaged shot noise discussed in the last subsection, the calculation of conductance fluctuation is different. This is because the shot noise contains one trace while the conductance fluctuation has two traces as can be seen below,
\begin{eqnarray}
\langle[\mathrm{Tr}(T)]^2\rangle = \langle\mathrm{Tr}[G^r\Gamma_L G^a\Gamma_R]\mathrm{Tr}[G^r\Gamma_L G^a\Gamma_R]\rangle.
\end{eqnarray}
If we still use the same idea as that of shot noise, we will find the calculation becomes more complicated because we can only write the above equation as
\begin{eqnarray}
\langle[\mathrm{Tr}(T)]^2\rangle = \sum_{ij}\langle(G^r\Gamma_L G^a\Gamma_R)_{ii}(G^r\Gamma_L G^a\Gamma_R)_{jj}\rangle\notag\\
=\sum_{ij}\langle\mathrm{Tr}[G^r\Gamma_L G^a\Gamma_R P^{ij} G^r\Gamma_L G^a\Gamma_R P^{ji}]\rangle,
\end{eqnarray}
in which the matrix $P^{ij}$ is the extremely sparse matrix with only one non-zero element, $(P^{ij})_{ij}=1$. It turns out that the mean value of $T^2$ will cost a factor of $N^2$ to the time scale as to evaluate shot noise. Even if we take into account from physics the propagation modes\cite{eric_unpublish}, $\Gamma_R = \sum_m|W_m\rangle\langle W_m|$, we still have 
\begin{eqnarray}
&&\langle[\mathrm{Tr}(T)]^2\rangle  \notag \\
=&&\sum_{mn}\langle\mathrm{Tr}[G^r\Gamma_L G^a S^{mn} G^r\Gamma_L G^a (S^{mn})^\dagger]\rangle,
\end{eqnarray}
where $|W_m\rangle$ represents the $m$th non-evanescent mode of right lead and $S^{mn}$ is defined as $|W_m\rangle\langle W_n|$. In this case, the factor of the computational cost is the square of the number of the non-evanescent modes, still difficult. However, in our direct expansion approach, we can get rid of this difficulty by taking the advantage of the property of $T_i$, Eq.(\ref{Ti_component}), see below.

\begin{figure}[htbp]
\includegraphics[height=0.9in,width=1.5in]{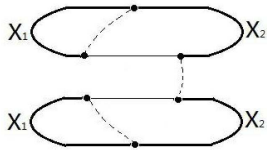}
\caption{One typical sixth-order diagram included in the evaluation of conductance fluctuation labeled as (12,21) with index contraction (i,ij,kj,k).}
\label{fluct_diagram_1221}
\end{figure}

As before, we substitute Eq.(\ref{Tmat_expansion}) into the above equation and expand it in terms of T-matrix. Here we take the term involving four T-matrices as an example, which is
\begin{eqnarray*}
&&\langle\mathrm{Tr}[G_e^rT^rG_e^r\Gamma_L G_e^aT^aG_e^a\Gamma_R]\mathrm{Tr}[G_e^rT^rG_e^r\Gamma_L G_e^aT^aG_e^a\Gamma_R]\rangle\\
&&=\langle\mathrm{Tr}[T^r X_1 T^a X_2]\mathrm{Tr}[T^r X_1 T^a X_2]\rangle.
\end{eqnarray*}
Up to the sixth-order in $T_i$, there are many diagrams with different ways of contraction for site indices. Now
considering a particular diagram (12,21) where the first two indices are in the first trace and the second two are in the second trace and a specific index contraction
$(i,ij,kj,k)$ as an example, Fig.(\ref{fluct_diagram_1221}),

\noindent whose contribution is
\begin{eqnarray}\label{fluct_connect}
&&\sum'_{ijk}\langle\mathrm{Tr}[T^r_i X_1 T_i^a G_e^a T_j^a X_2]\mathrm{Tr}[T_k^r G_e^r T_j^r X_1 T_k^a X_2]\rangle \nonumber \\
=&&\sum'_{ijk}(X_1)_{ii}(G_e^a)_{ij}(X_2)_{ji}(G_e^r)_{kj}(X_1)_{jk}(X_2)_{kk} \nonumber \\
\times &&\langle\tau_i^r\tau_i^a\rangle\langle\tau_j^r\tau_j^a\rangle\langle\tau_k^r\tau_k^a\rangle,
\end{eqnarray}
where we have used Eq.(\ref{Ti_component}) to deal with two traces. In order to calculate the conductance
fluctuation, we need to evaluate both $\langle\mathrm{Tr}[T]^2\rangle$ and $\langle\mathrm{Tr}[T]\rangle$. We
notice that $\langle\mathrm{Tr}[T]\rangle$ can be calculated accurately using NVC while $\langle\mathrm{Tr}[T]^2\rangle$
can only be obtained in direct expansion. To make sure the accuracy of conductance fluctuation, we have to treat these
two terms on the equal footing and use direct expansion on both terms. As an example, if we expand $\langle T^2\rangle$ to sixth order but still use CPA+NVC to evaluate $\langle T\rangle$, the fluctuation obtained is not very accurate. In Fig.\ref{fig1}(d), at $w=0.1$, we get $\langle T^2\rangle=394.3527$ from sixth-order expansion and $\langle T\rangle=19.8583$ from CPA+NVC, the fluctuation evaluated from these results is then larger than $0.02$, which deviates from the
exact result $0.0125$ quite a lot. However, our sixth-order cumulant expansion directly on fluctuation gives the result $0.0130$, better agreement compared with the exact one.

Actually, in order to get the conductance fluctuation, a better way is to do cumulant expansion, which discards all the ``disconnected diagrams" \cite{Kubo_cumulant}. The advantages of such ``cumulant expansion" include the following separate aspects: 1. We can directly attack the fluctuation instead of expand both $\langle T^2\rangle$ and $\langle T\rangle$, so the computational cost is reduced to nearly a half. 2. In this way we can naturally evaluate $\langle T^2\rangle$ and $\langle T\rangle$ on the same footing without to evaluate either of them, and also avoid the error stated in the above paragraph. 3. This cumulant expansion only include connected diagrams, making the physical meaning more clear because that the connected diagrams only contributes to $\langle T\rangle\langle T\rangle$, which is never needed when we concentrate on the conductance fluctuation.
In our case, in one specific index combination, if the indices in the first trace do not connect to those in the second trace, then it is a disconnected terms. For example, in the decomposition (11,22), one disconnected term is $(i,i,jk,kj)$, while $(i,j,ki,kj)$ is a connected term. To a certain order, the sum of all the connected terms give the square of the conductance fluctuation.



\section{numerical results}\label{numerical_results}

\begin{figure}[htbp]
 \includegraphics[height=1.2in]{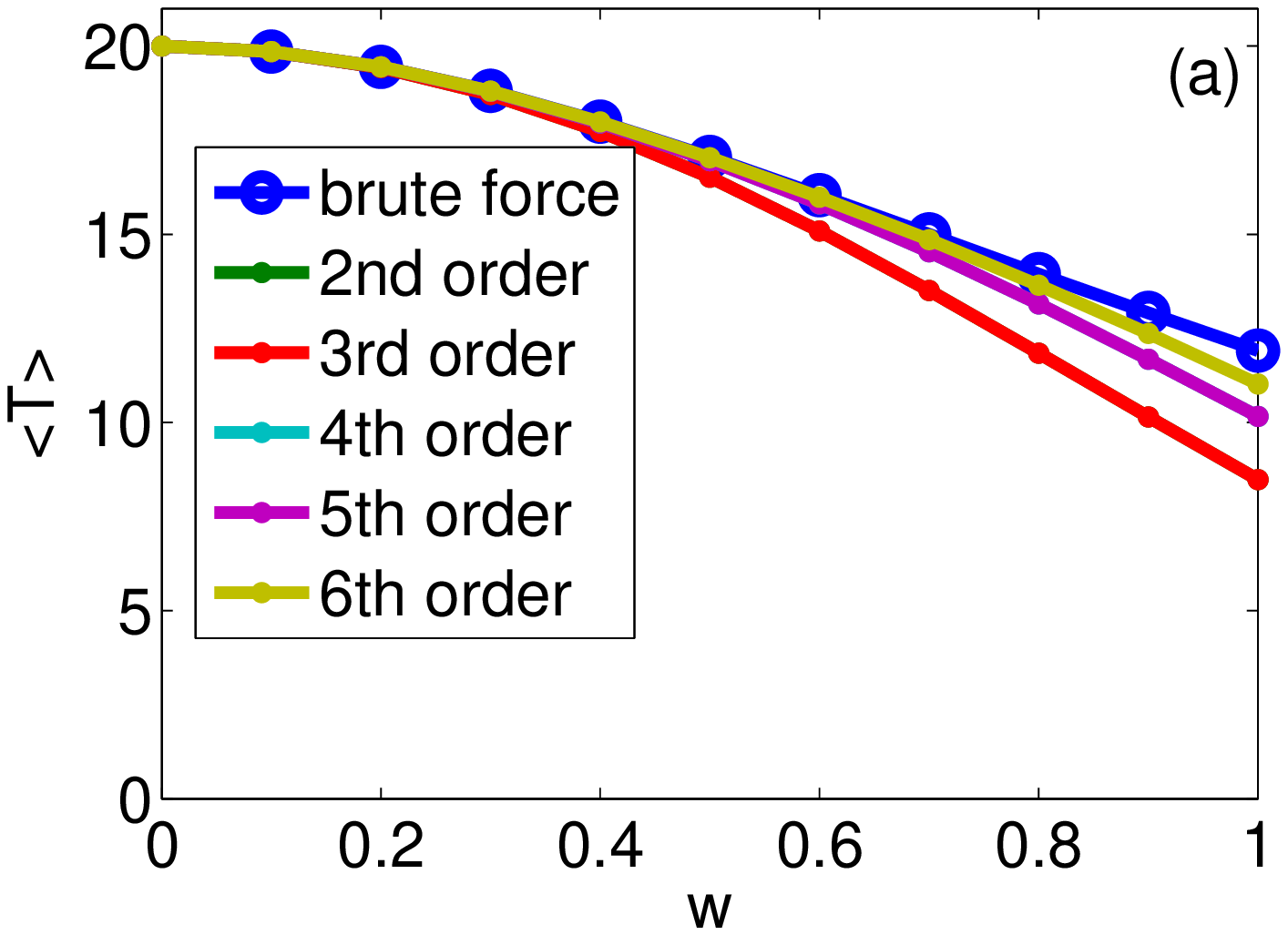}%
 \includegraphics[height=1.2in]{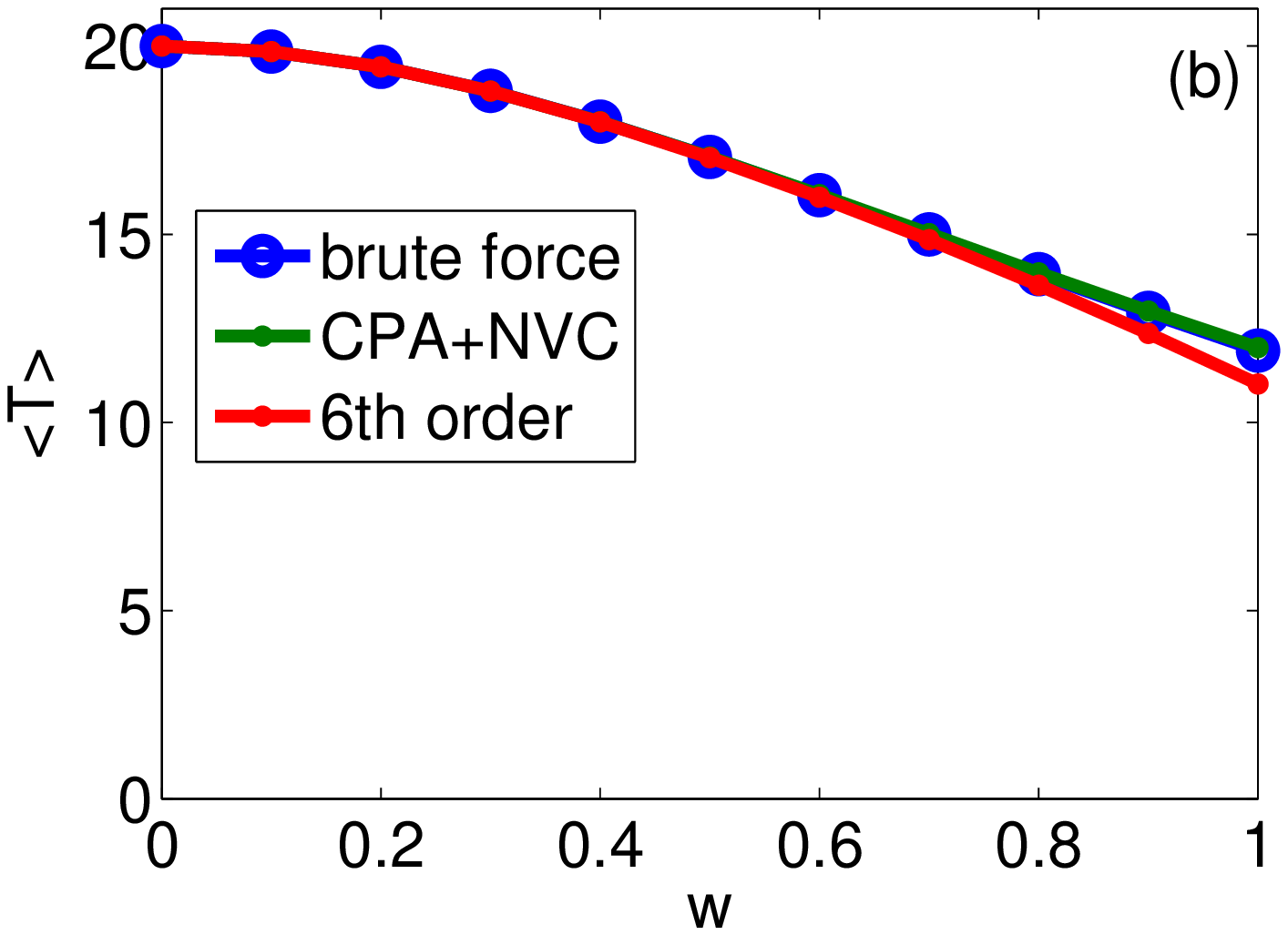}\\%
 \includegraphics[height=1.2in]{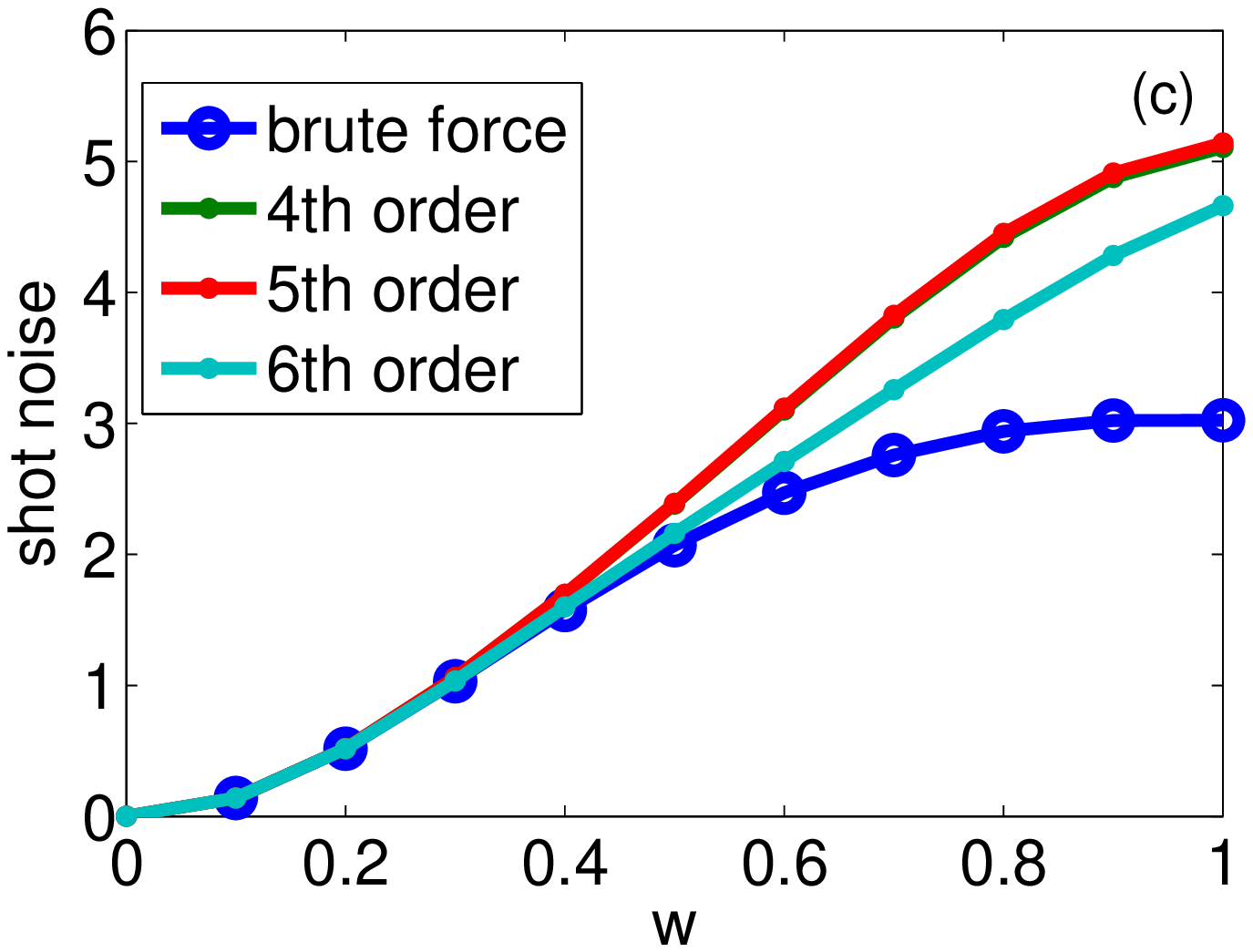}%
 \includegraphics[height=1.2in]{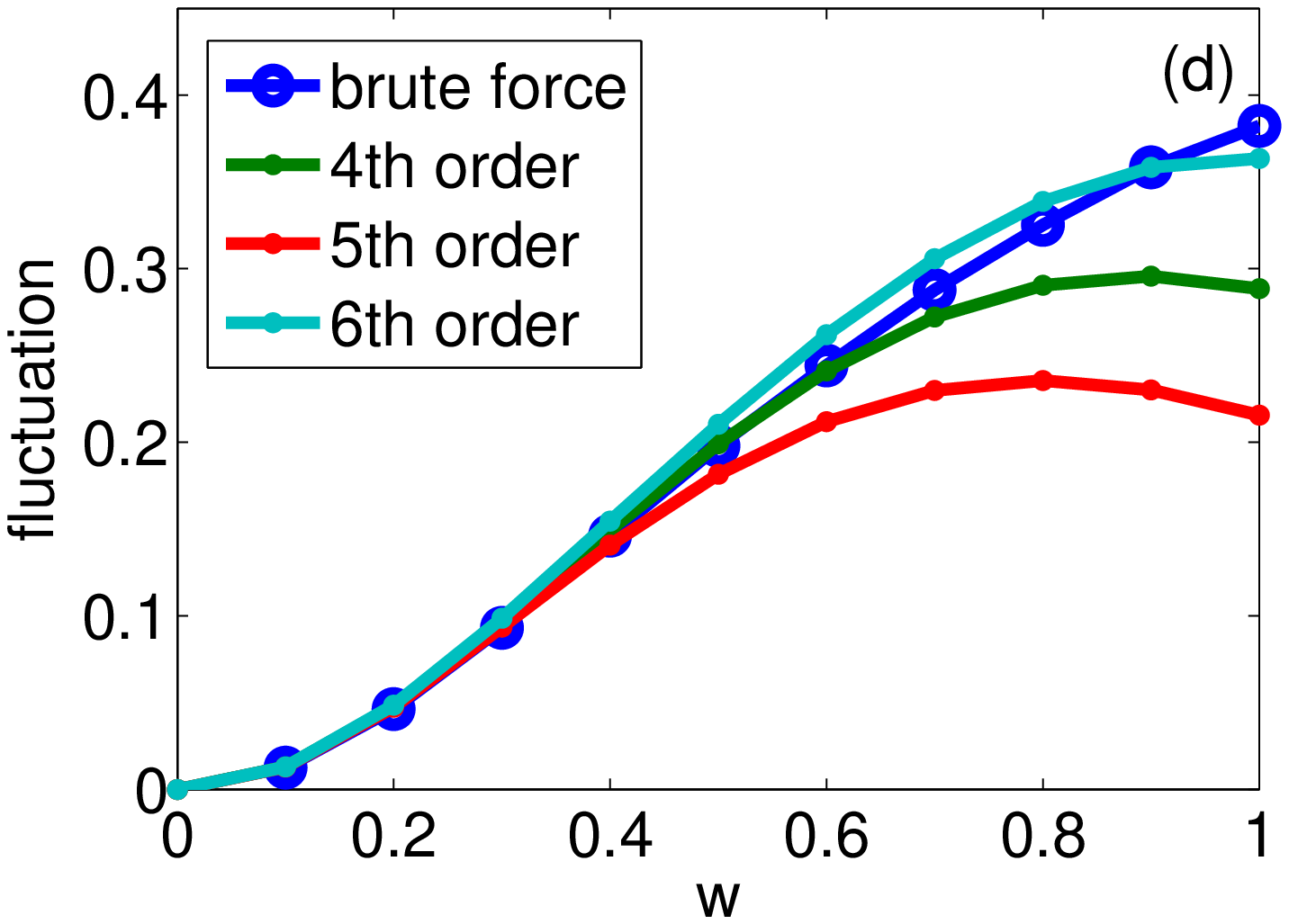}\\%
\caption{Square lattice of size $40 \times 40$ with Anderson disorder and fixed energy $E=2$.
(a) Conductance, direct expansion at different orders vs brute force.
(b) Conductance, direct expansion up to 6th order vs brute force vs NVC.
(c) Averaged shot noise, direct expansion at different orders vs brute force.
(d)  Conductance fluctuation, direct expansion at different orders vs brute force.}\label{fig1}%
\end{figure}

\begin{figure}[htbp]
\includegraphics[height=1.16in]{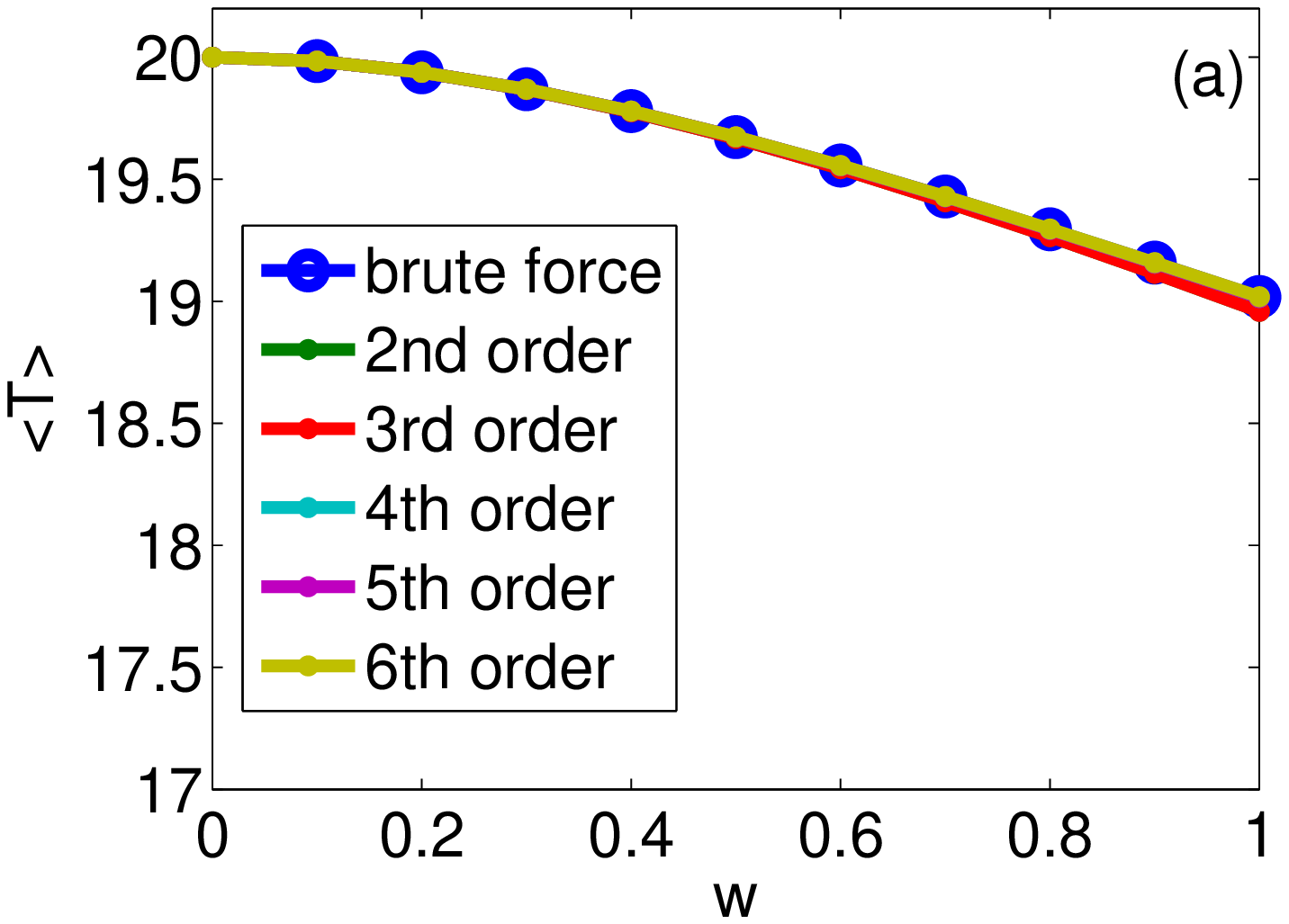}\label{fig2a}
\includegraphics[height=1.16in]{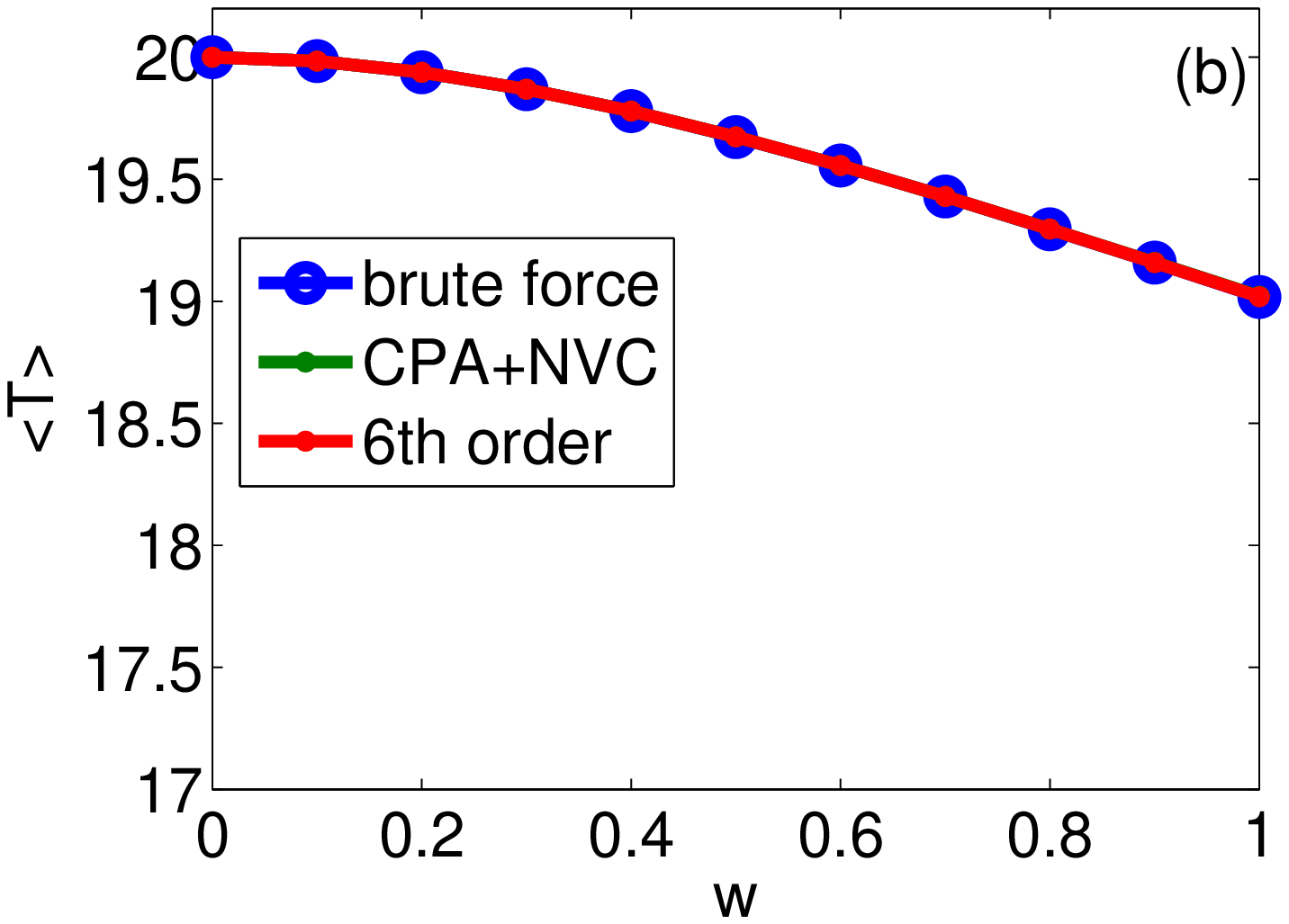}\label{fig2b}\\
\includegraphics[height=1.16in]{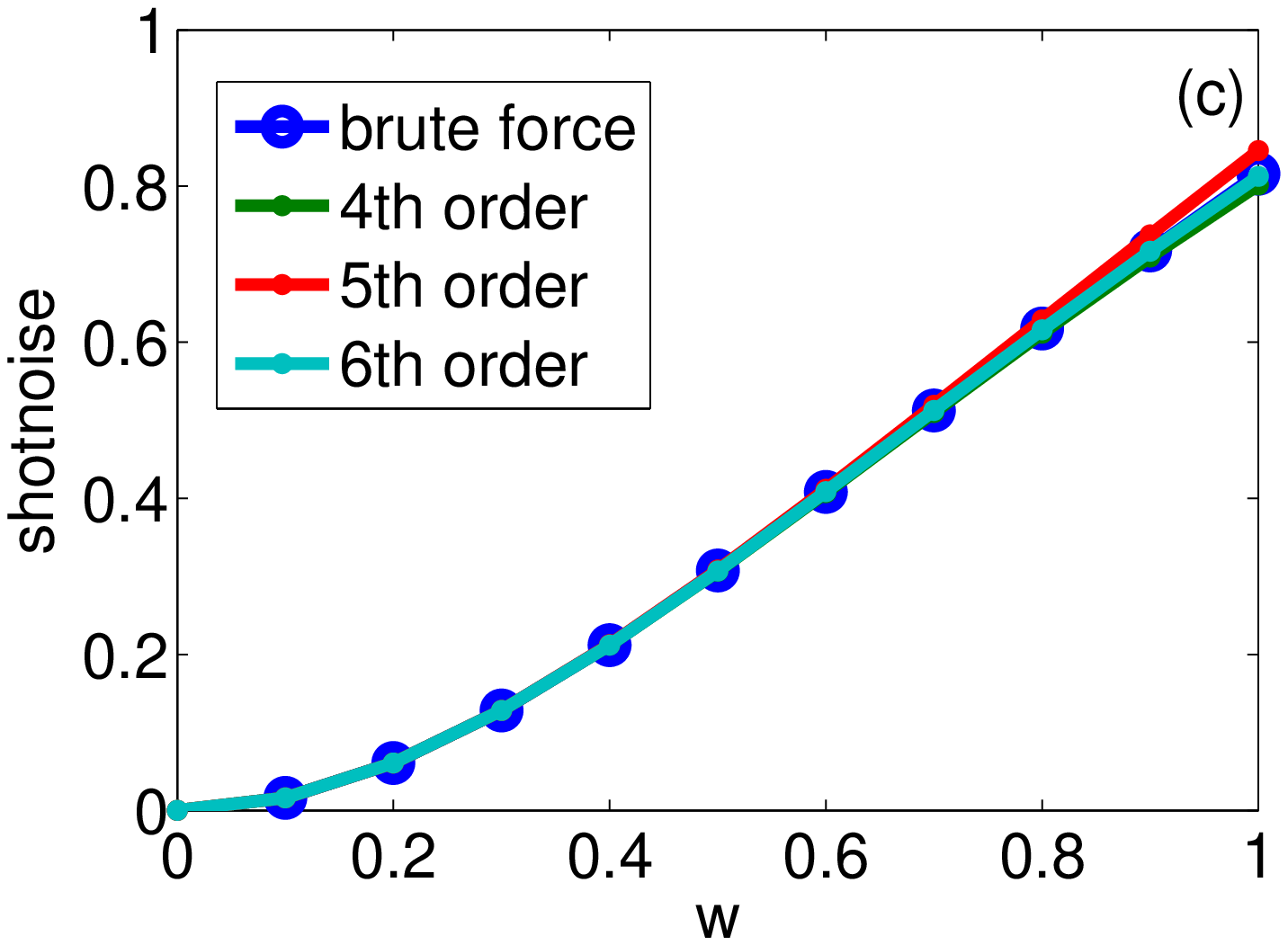}\label{fig2c}
\includegraphics[height=1.16in]{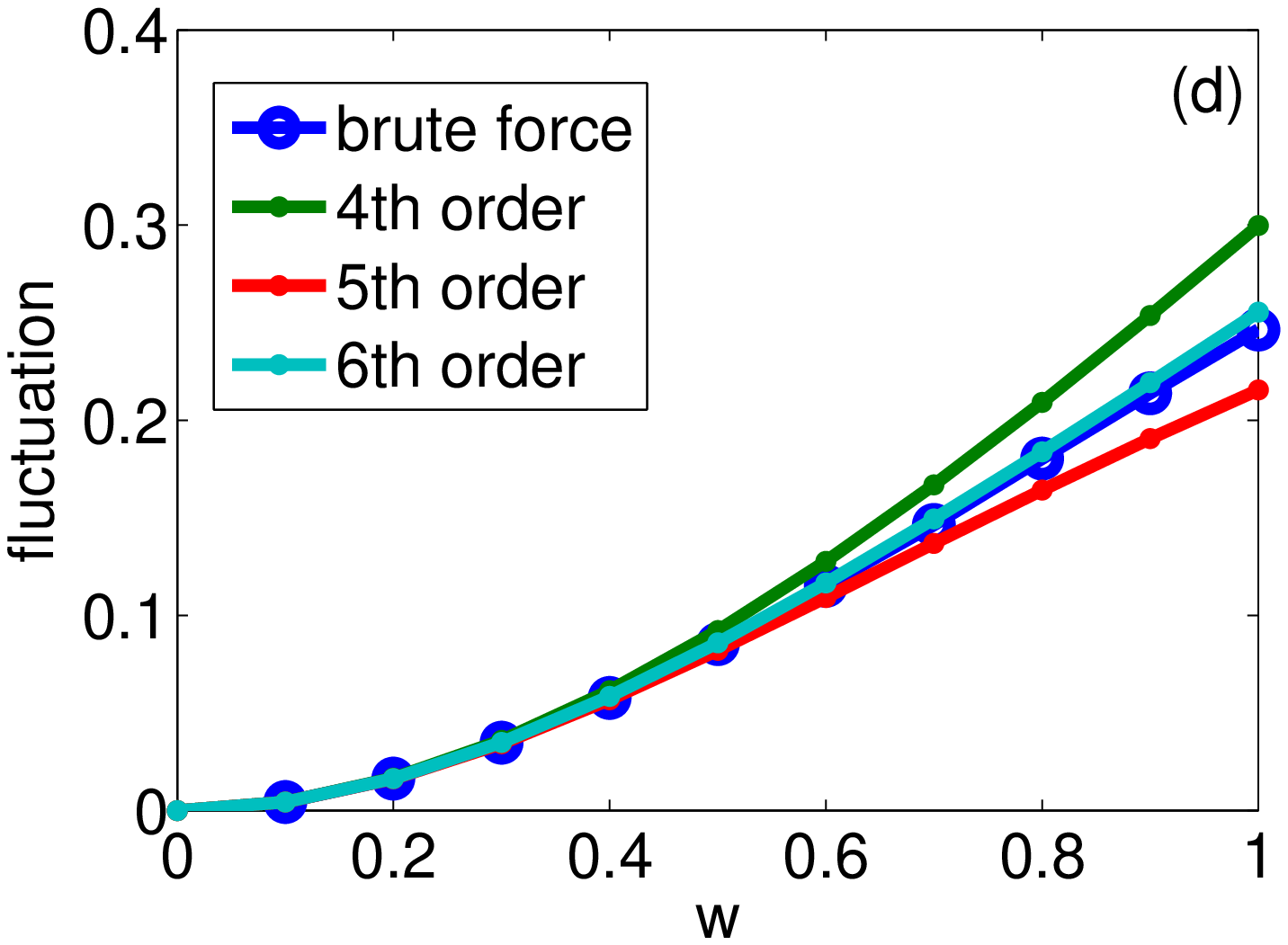}\label{fig2d}\\
 \caption{Square lattice of size $40 \times 40$ with $1\%$ doping concentration and fixed energy $E=2$.
 (a) Conductance, direct expansion at different orders vs brute force.
 (b) Conductance, direct expansion up to 6th order vs brute force vs NVC.
 (c) Averaged shot noise, direct expansion at different orders vs brute force.
 (d)  Conductance fluctuation, direct expansion at different orders vs brute force.}\label{fig2}%
\end{figure}

Before we show the numerical results we wish to mention the computational cost of our approach. As we can see from
the algorithm, first we need to generate all the topologically inequivalent diagrams of $T_i$ up to certain order.
Secondly we have to generate all the possible index contractions for a given diagram. As we go to higher order, both
number of $T_i$ and the number of contractions for each $T_i$ grow exponentially. Note that due to the CPA condition,
Eq.(\ref{cpa_condition}), a diagram does not contribute if an index appears only once. Hence each index has to
appear at least twice in the summation. Thus, up to the $n$th order, the largest number of different indices in the
summation is $\lfloor n/2\rfloor$ which dominates the computational cost. In general an additional index will cost
about $N$ times computational time, with $N$ being the number of atoms. For this reason, although we have generated
all the diagrams up to the 8th order in $T_i$, we can only apply our approach to a small sized system such as
a 10-by-10 system in 2D in a reasonable amount of CPU time. In this paper, we apply our formalism
to 40-by-40 and 60-by-60 systems in 2D up to 6th order in $T_i$. Below we show the results of conductance,
shot noise, and conductance fluctuations where we consider Anderson disorder with different disorder strength and
doping with low (1\%) and high (10\%) doping concentrations.

\begin{figure}[htbp]
 \includegraphics[height=1.18in]{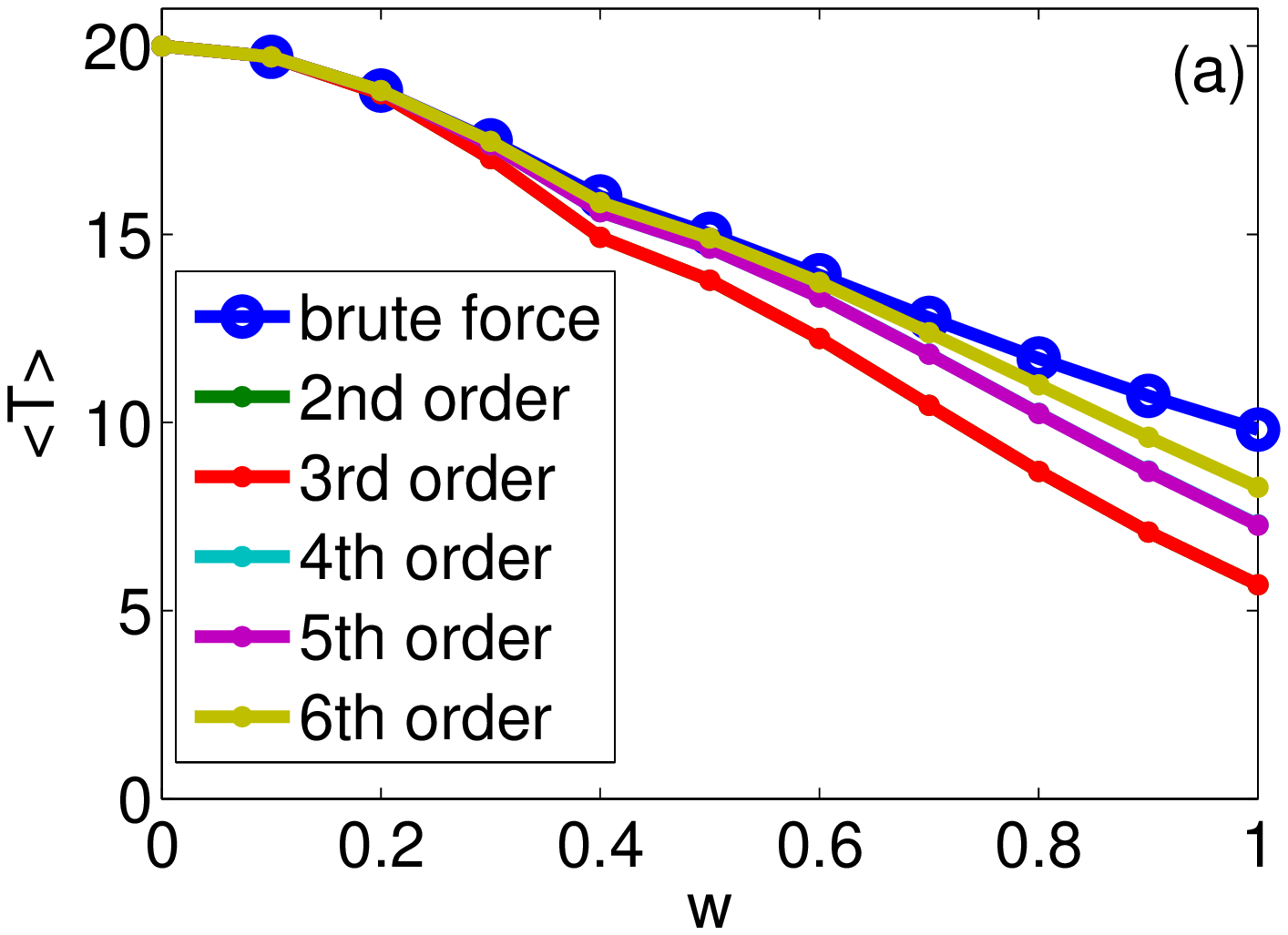}
 \includegraphics[height=1.18in]{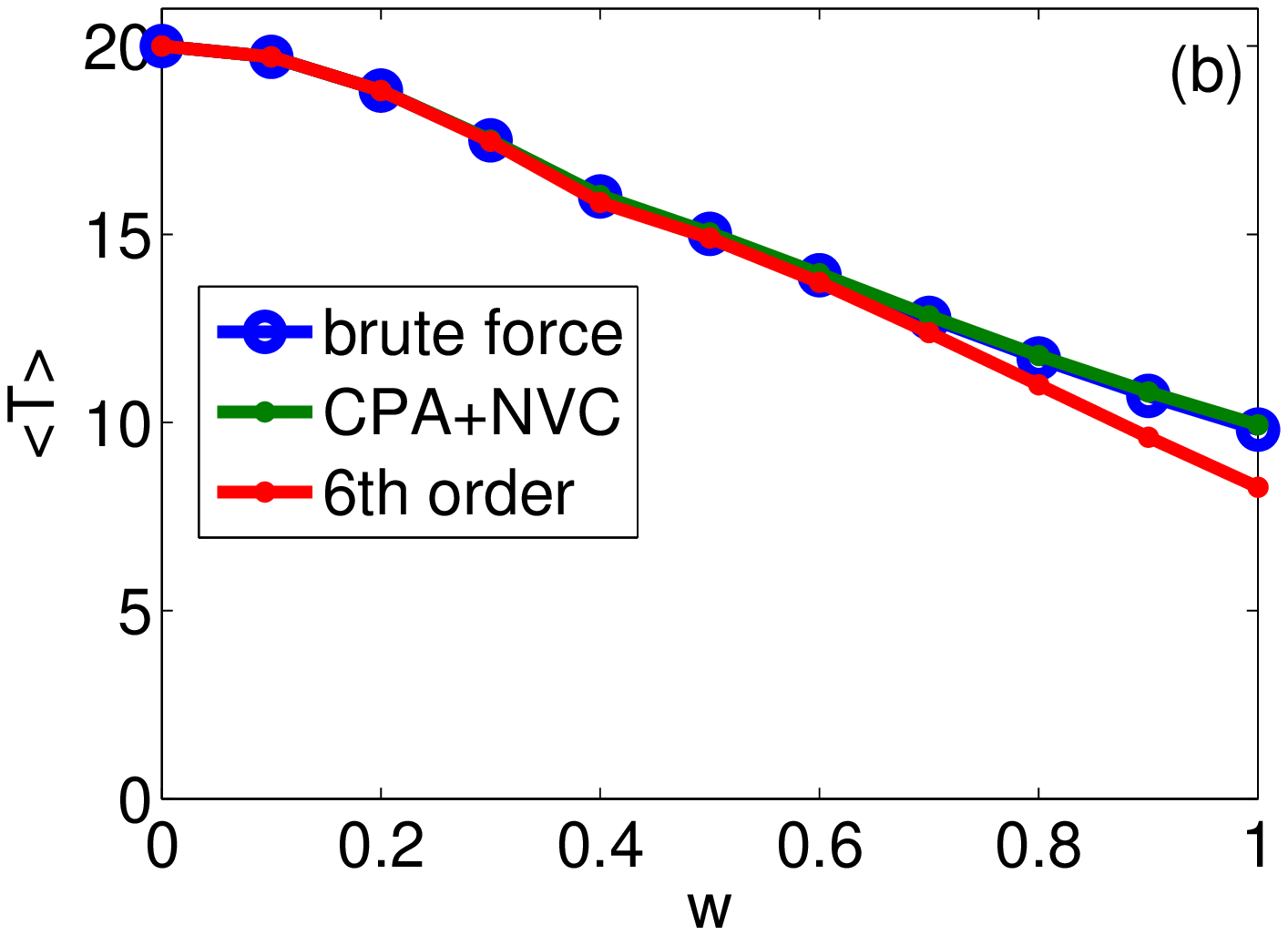}\\
 \includegraphics[height=1.18in]{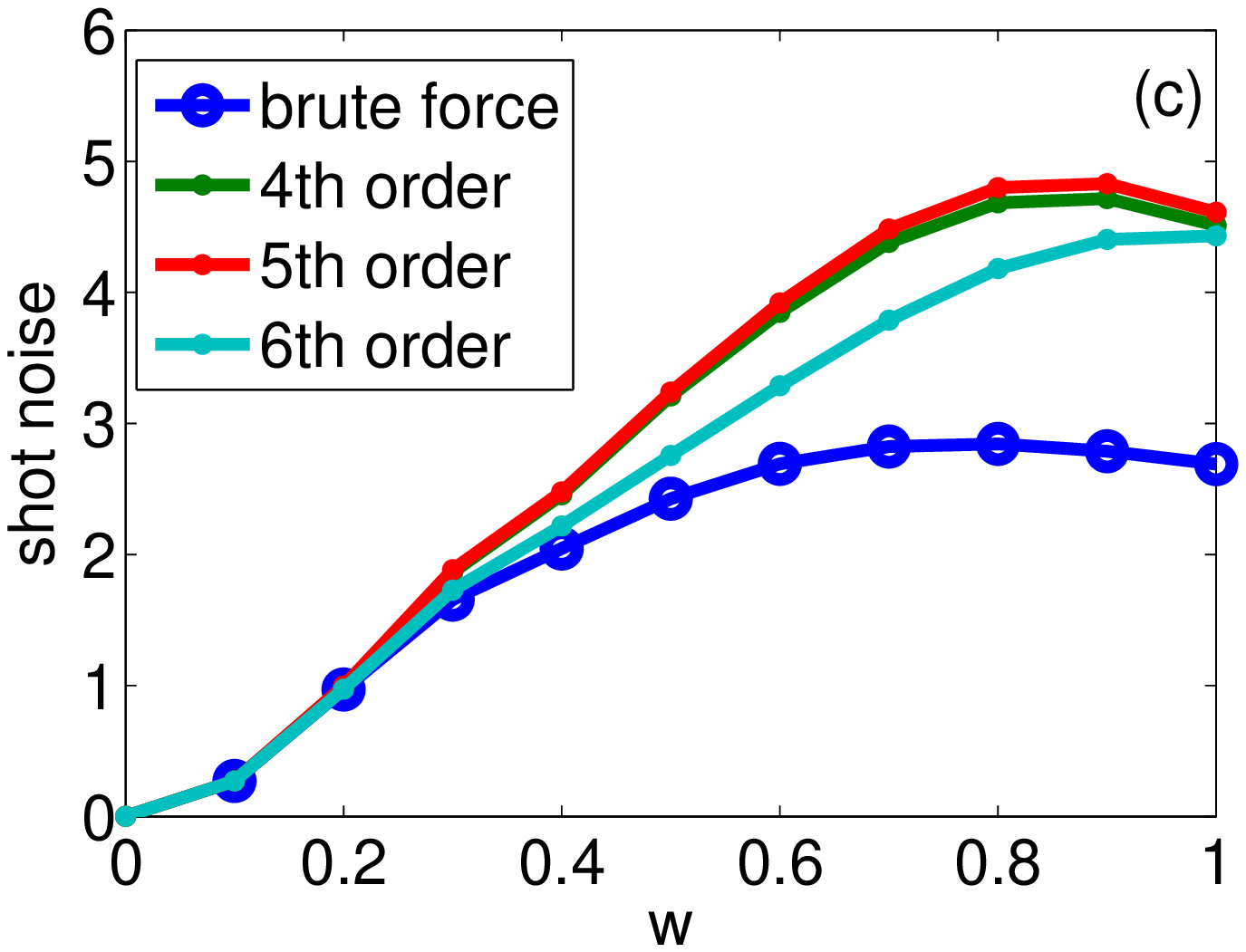}
 \includegraphics[height=1.18in]{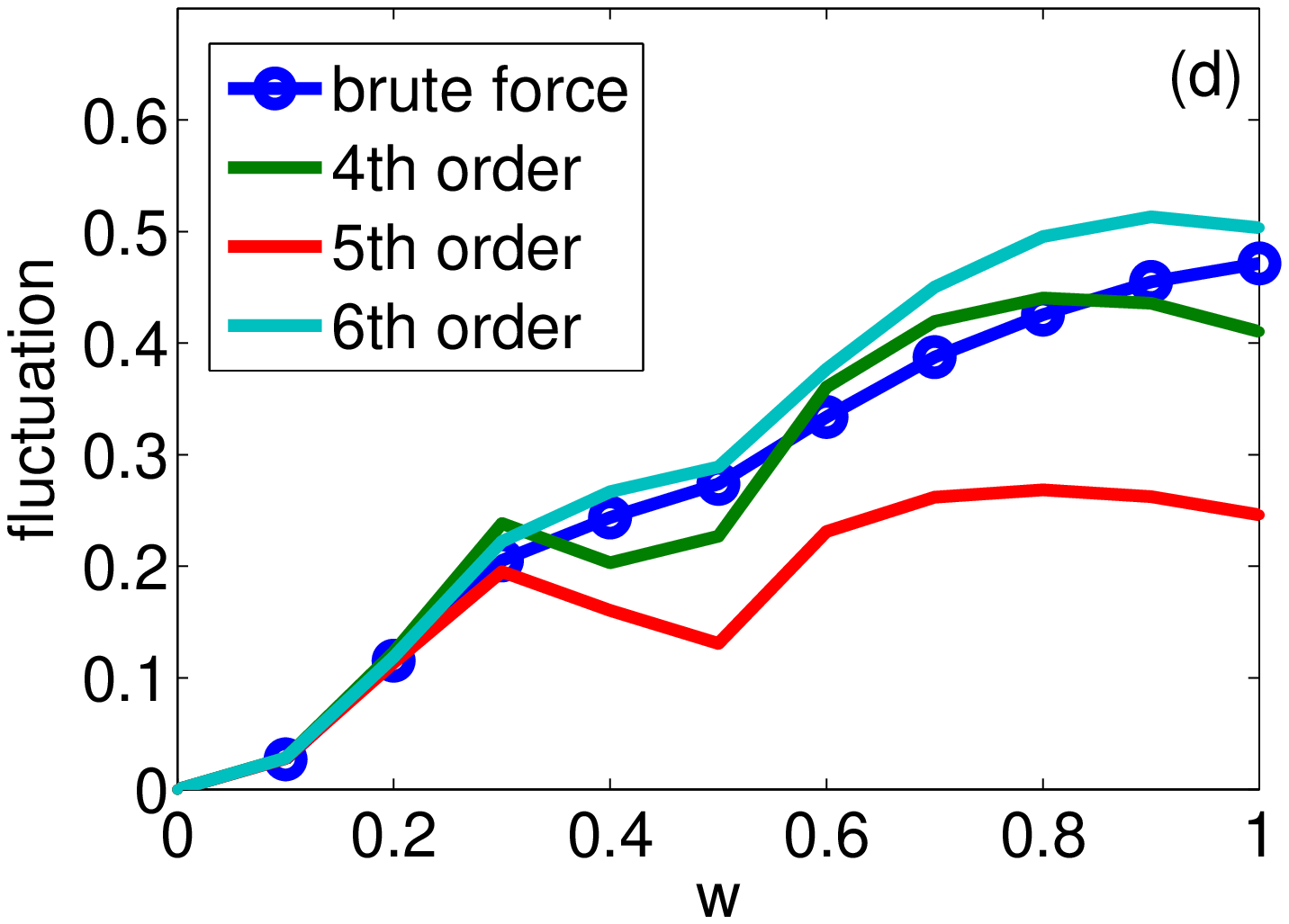}\\
 \caption{Square lattice of size $40 \times 40$ with $20\%$ doping concentration and fixed energy $E=2$.
 (a) Conductance, direct expansion at different orders vs brute force.
 (b) Conductance, direct expansion up to 6th order vs brute force vs NVC.
 (c) Averaged shot noise, direct expansion at different orders vs brute force.
 (d) Conductance fluctuation, direct expansion at different orders vs brute force.}\label{fig3}%
\end{figure}

\begin{figure}[htbp]
 \includegraphics[height=1.19in]{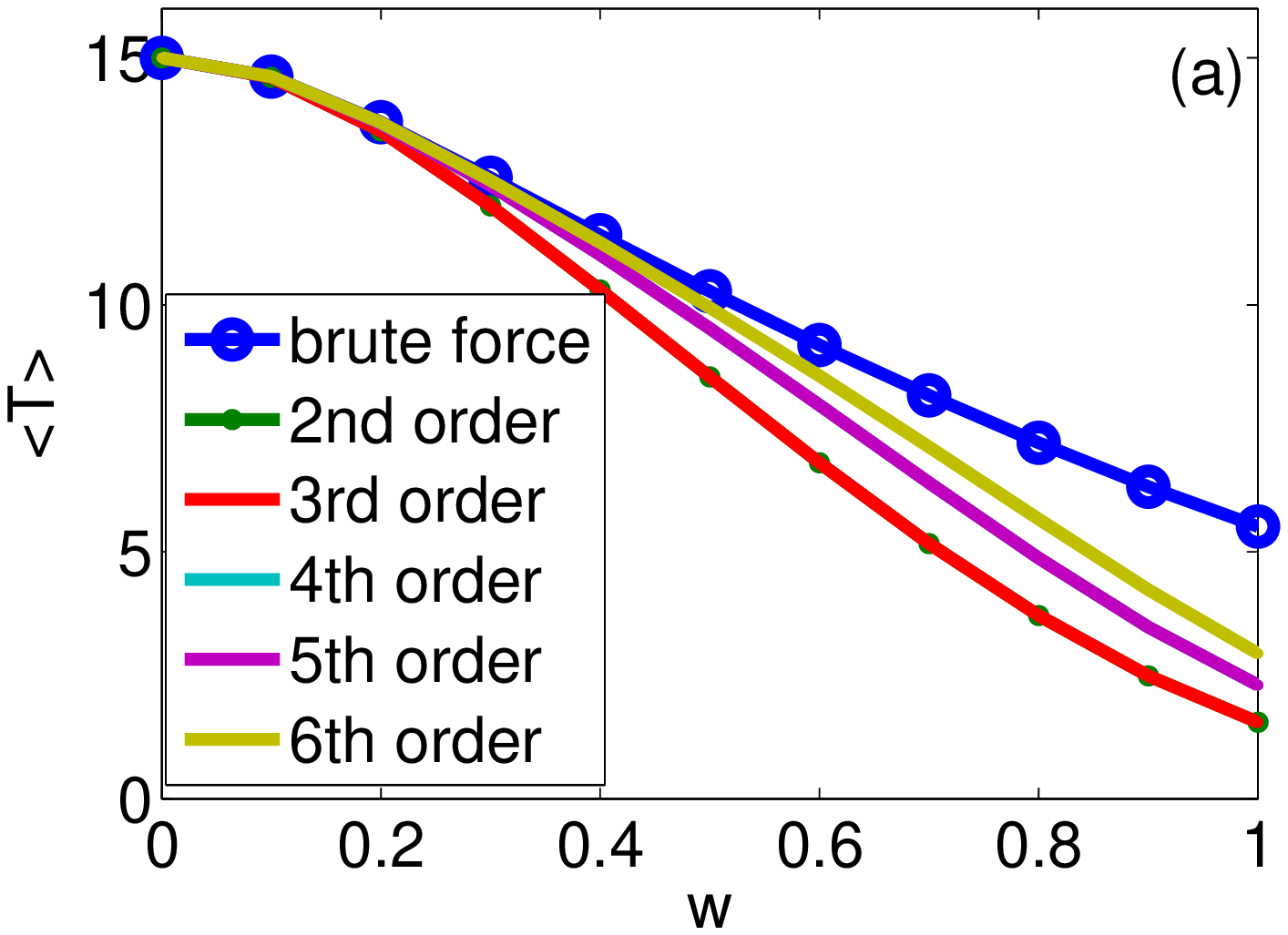}
 \includegraphics[height=1.19in]{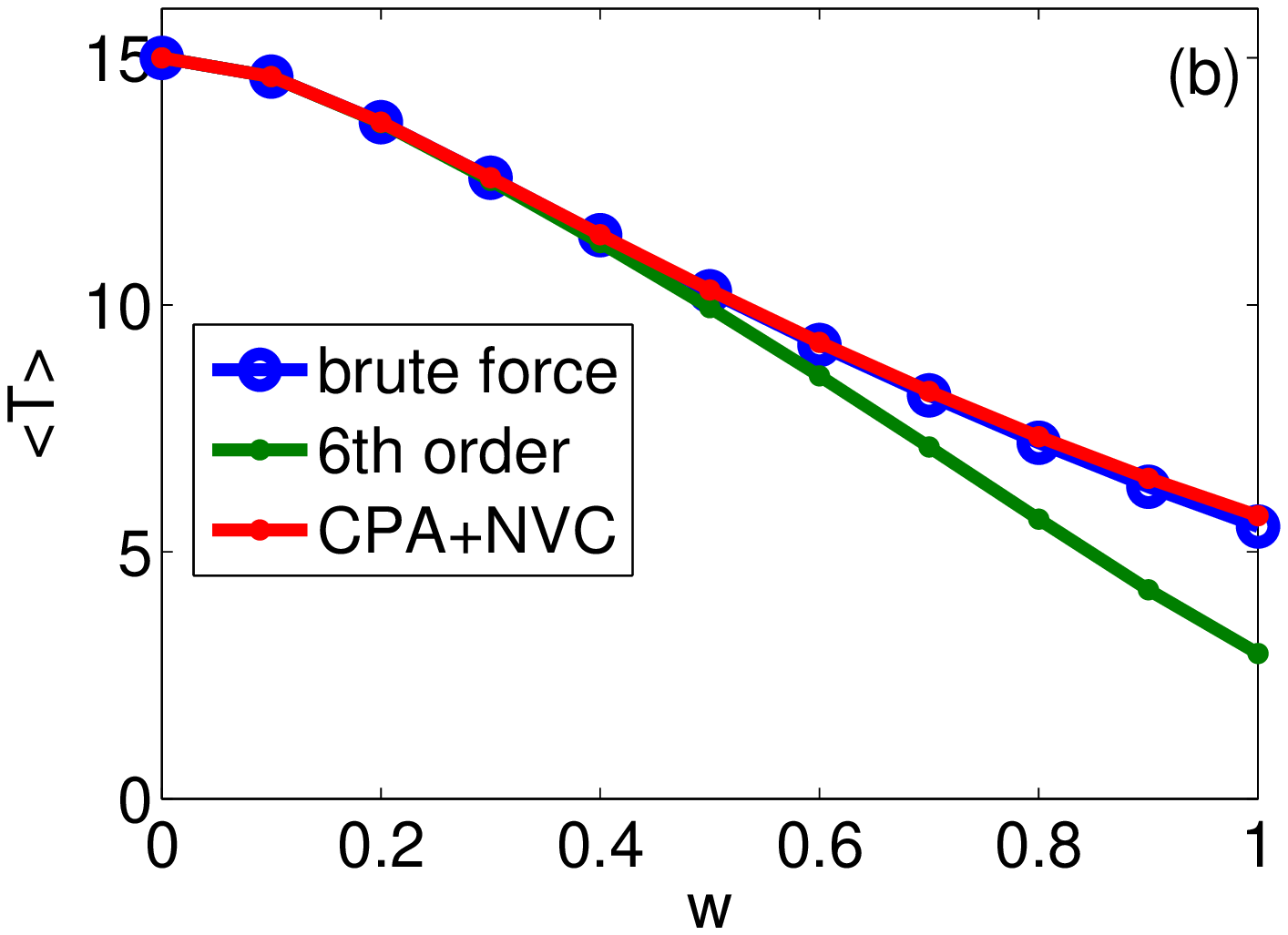}\\
 \includegraphics[height=1.19in]{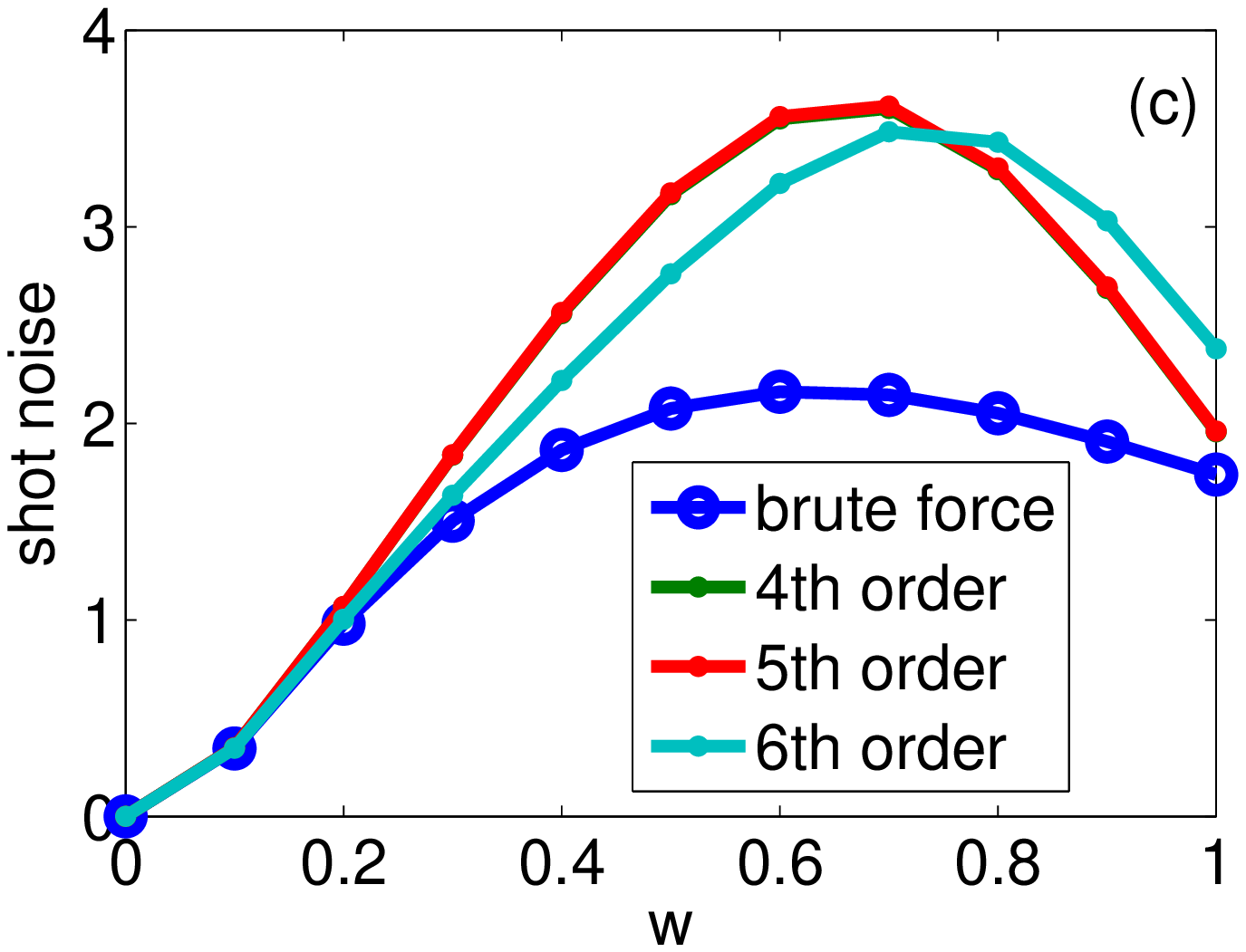}
 \includegraphics[height=1.19in]{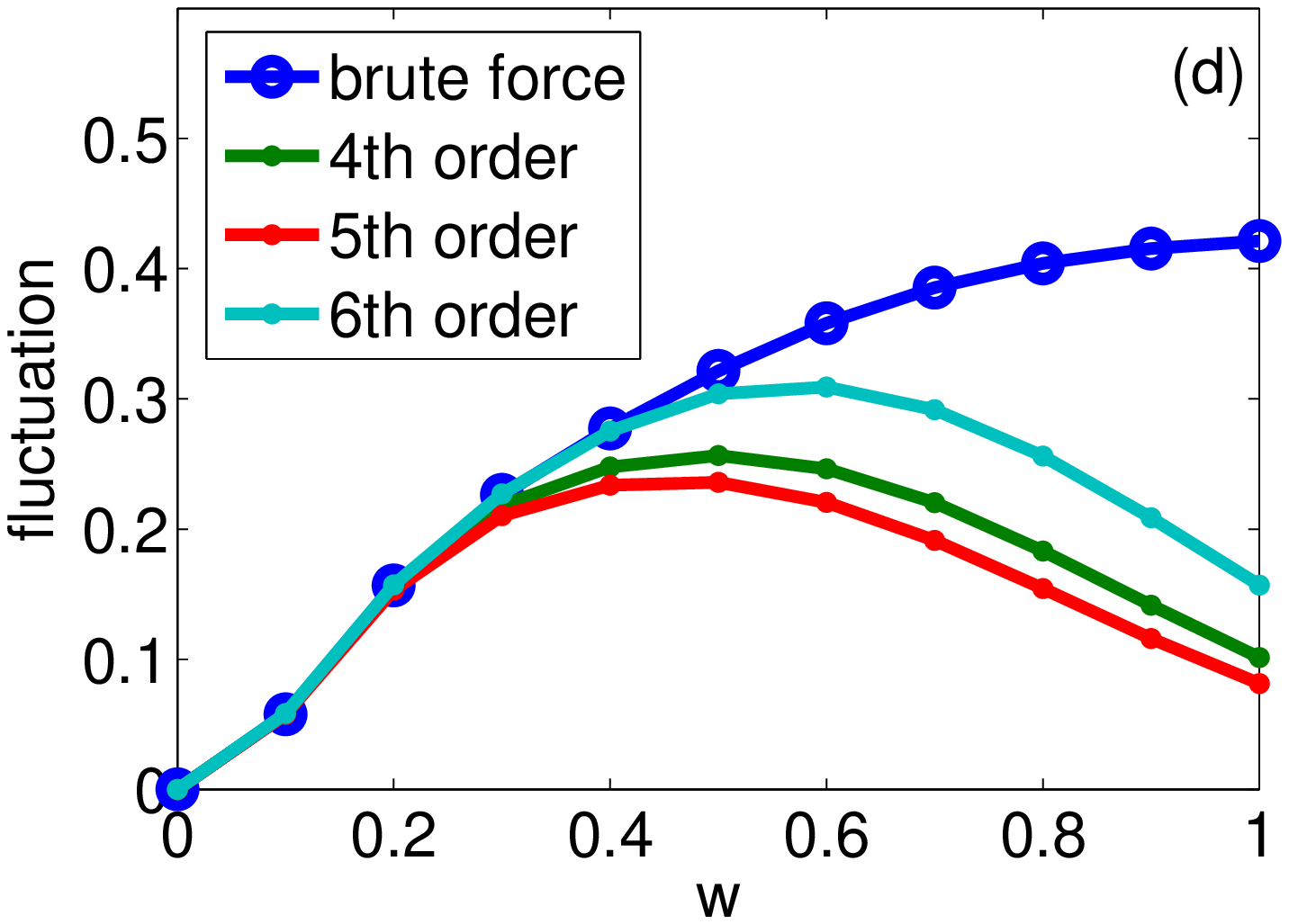}\\
 \caption{20 layered graphene, Anderson disorder, $E=0.55$. (a) Conductance, direct expansion up to different order vs brute force. (b) Conductance, direct expansion up to 6th order vs brute force vs NVC. (c) Averaged shot noise, direct expansion up to different order vs brute force. (d)  Conductance fluctuation, direct expansion up to different order vs brute force.}\label{fig9}%
\end{figure}

\begin{figure}[htbp]
 \includegraphics[height=1.19in]{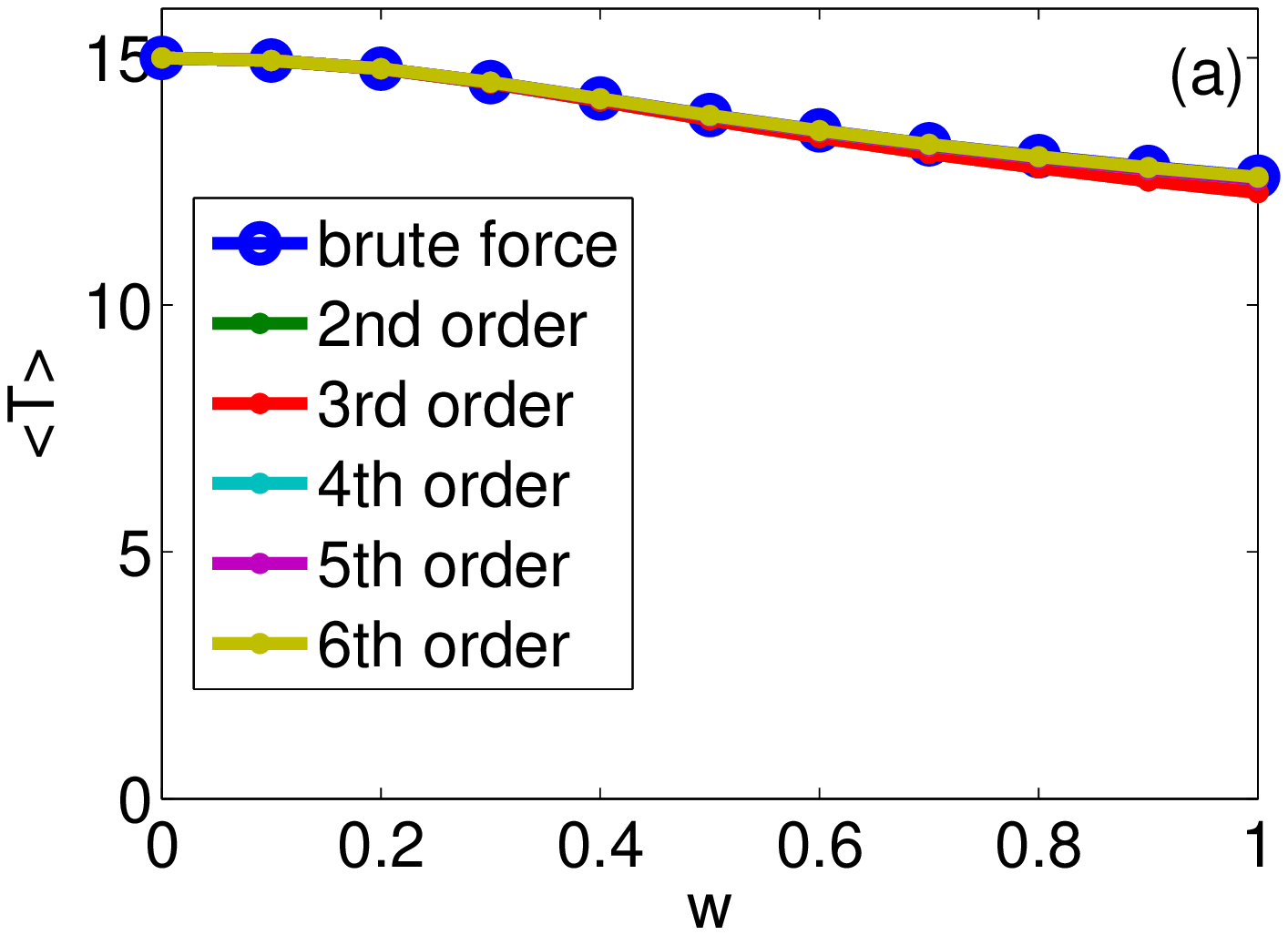}
 \includegraphics[height=1.19in]{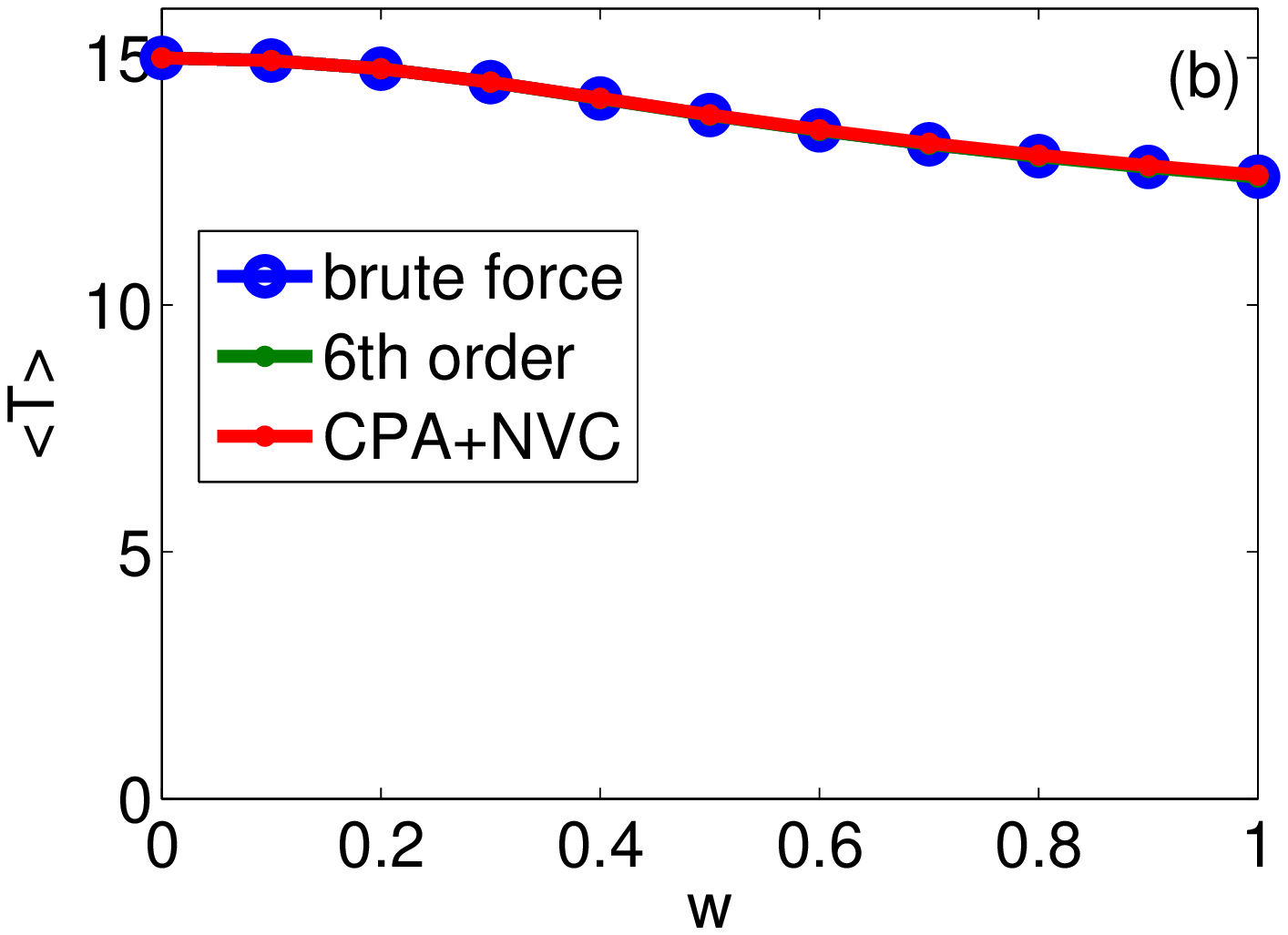}\\
 \includegraphics[height=1.19in]{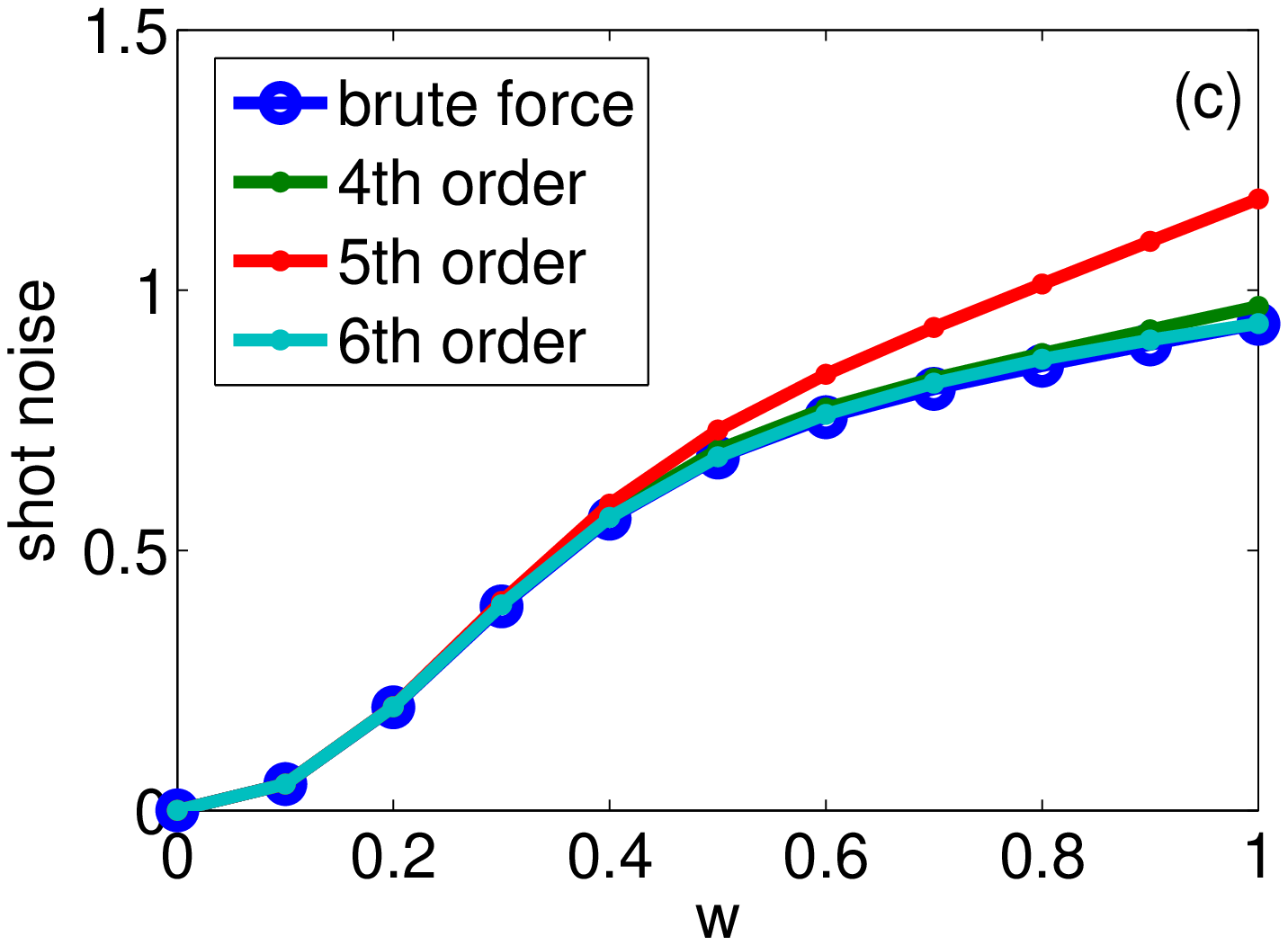}
 \includegraphics[height=1.19in]{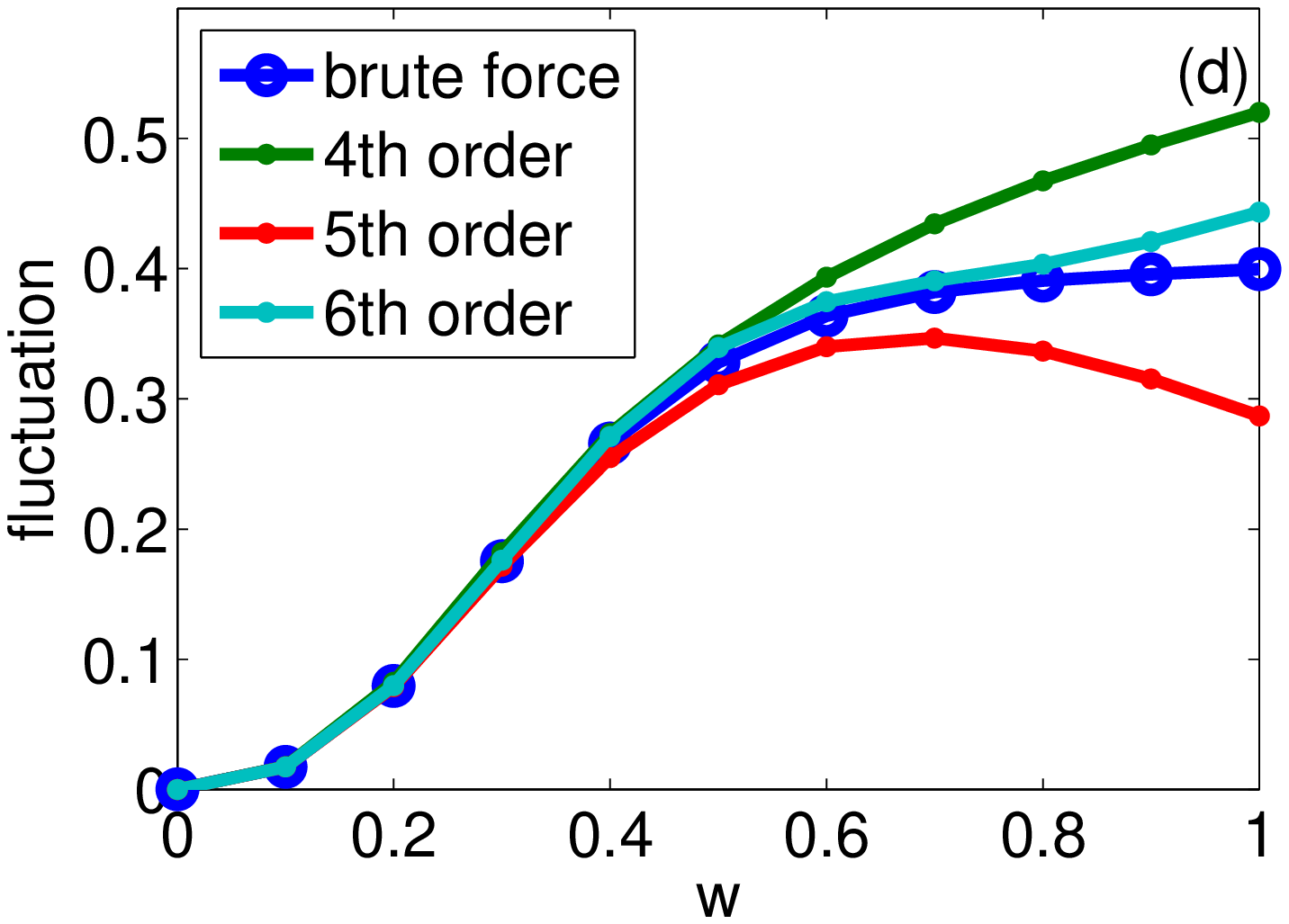}\\
 \caption{20 layered graphene, 1\% doping , $E=0.55$. (a) Conductance, direct expansion up to different order vs brute force. (b) Conductance, direct expansion up to 6th order vs brute force vs NVC. (c) Averaged shot noise, direct expansion up to different order vs brute force. (d)  Conductance fluctuation, direct expansion up to different order vs brute force.}\label{fig10}%
\end{figure}

\begin{figure}[htbp]
 \includegraphics[height=1.19in]{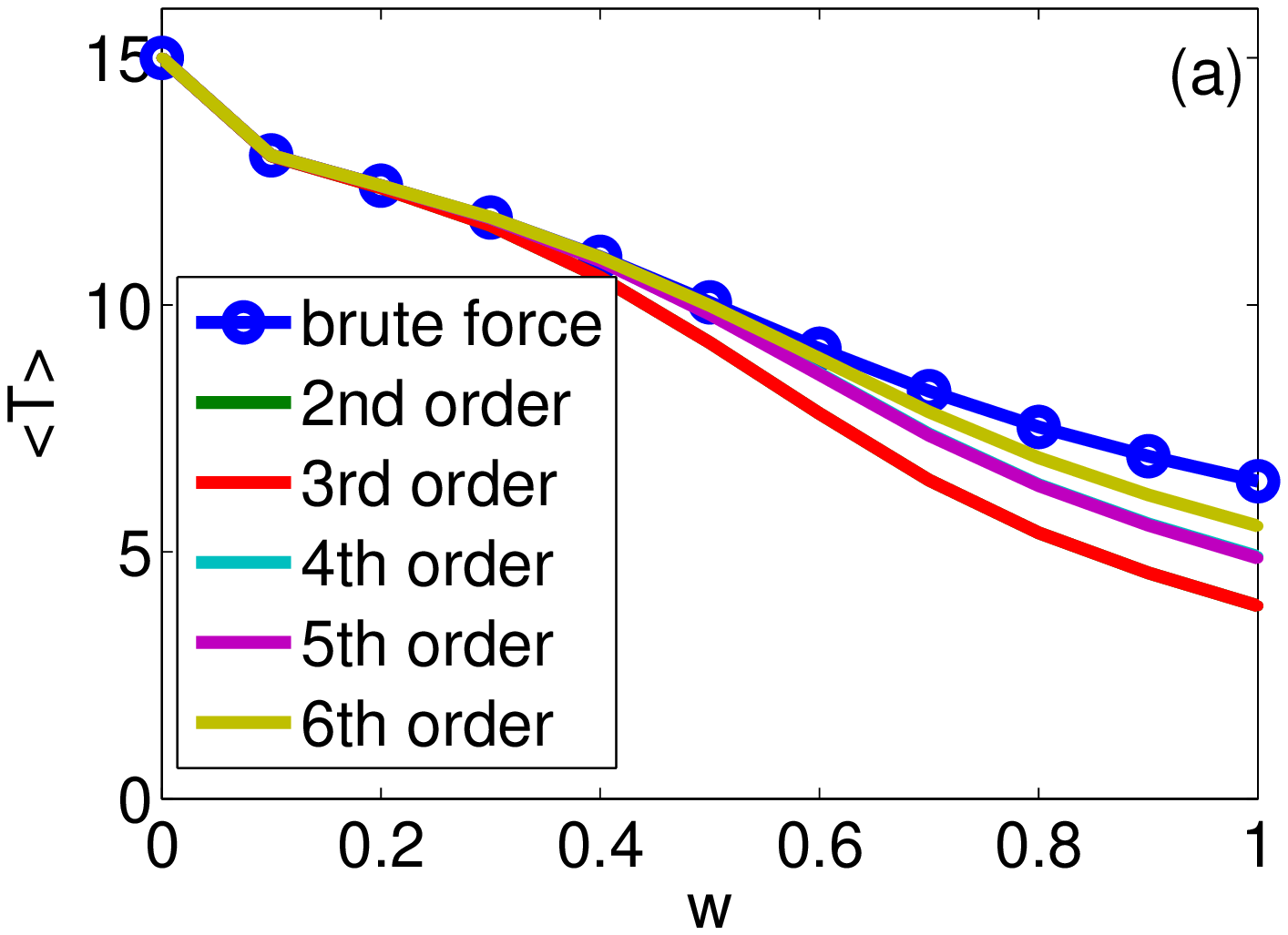}
 \includegraphics[height=1.19in]{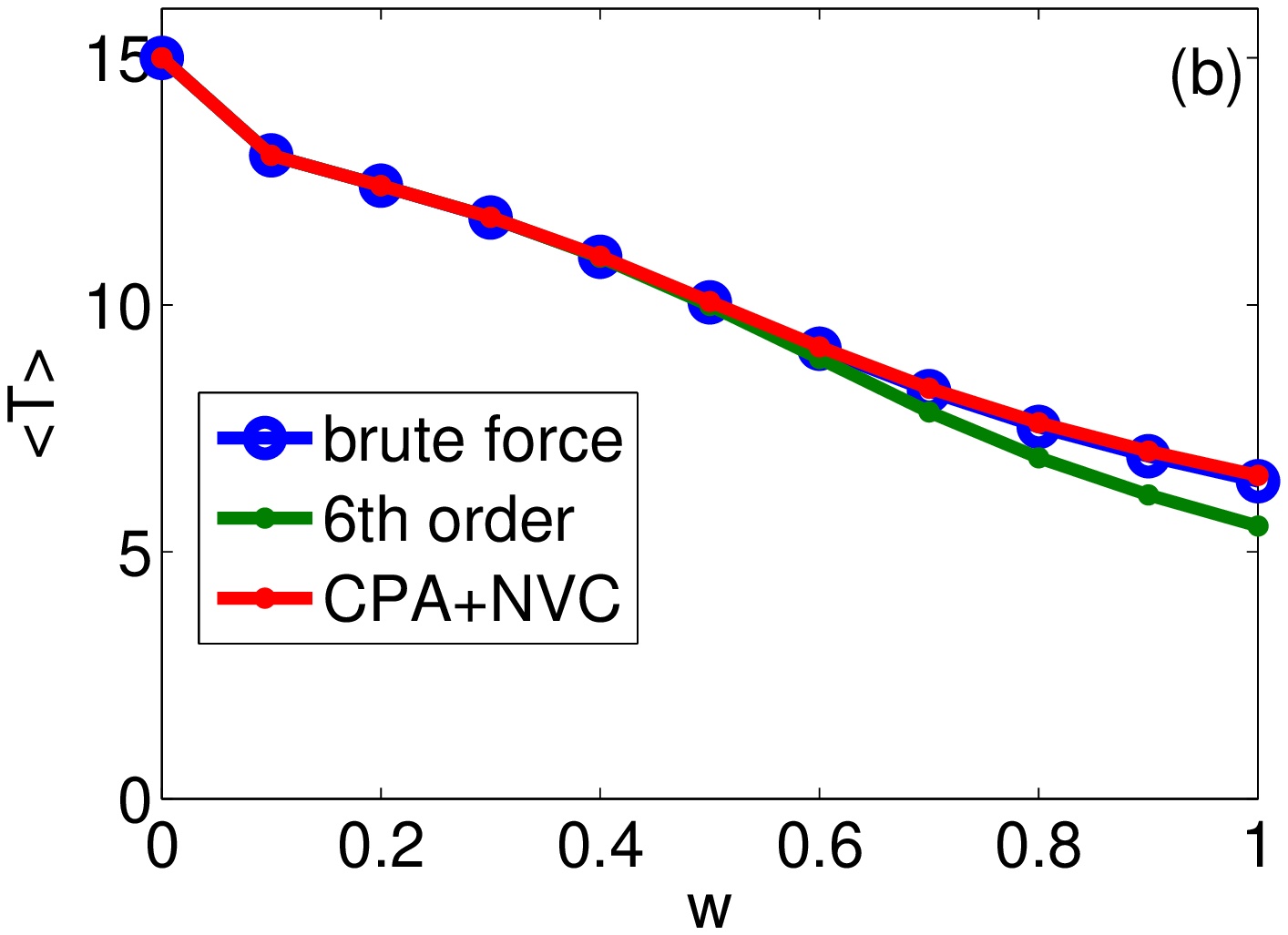}\\
 \includegraphics[height=1.19in]{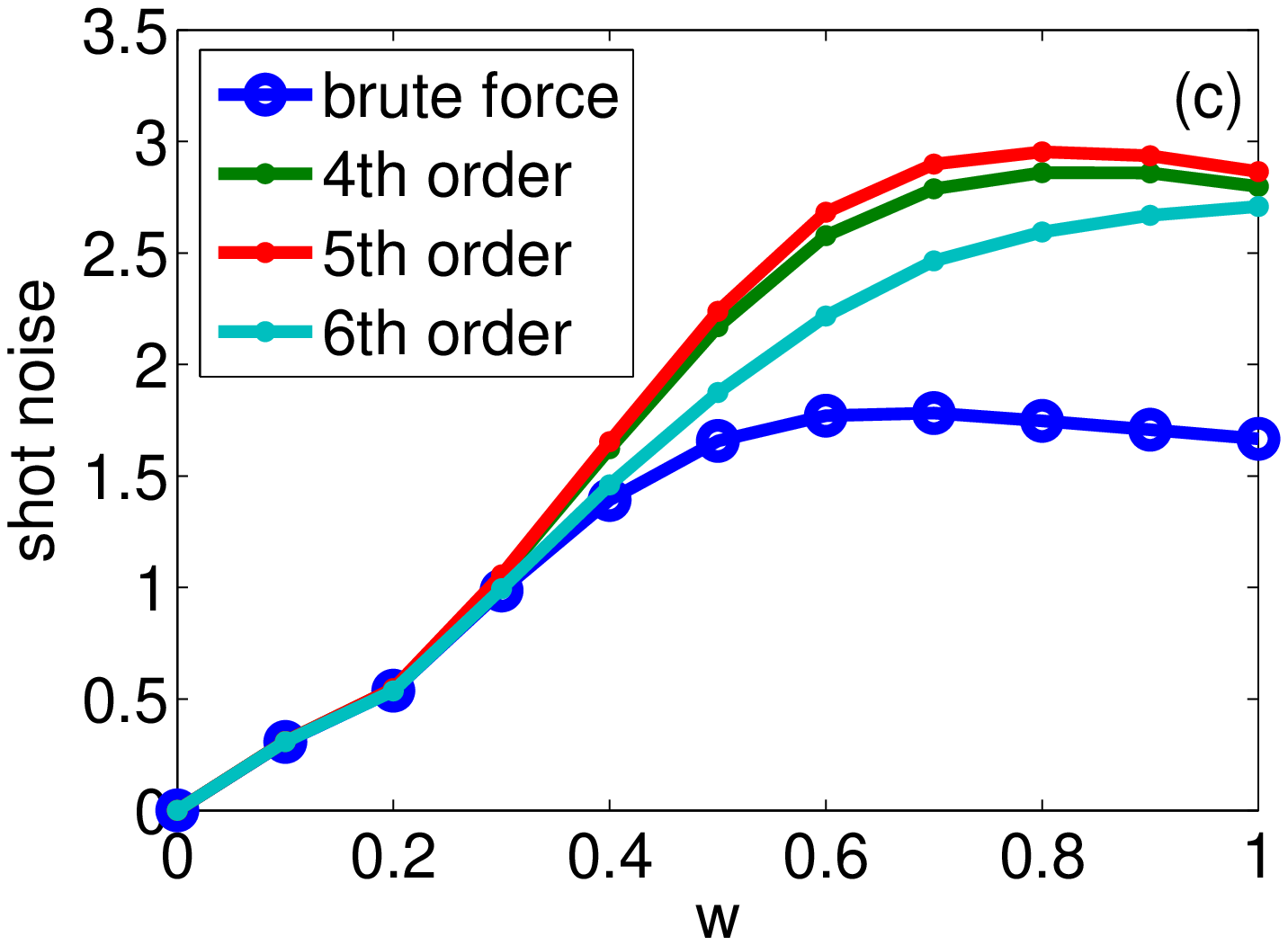}
 \includegraphics[height=1.19in]{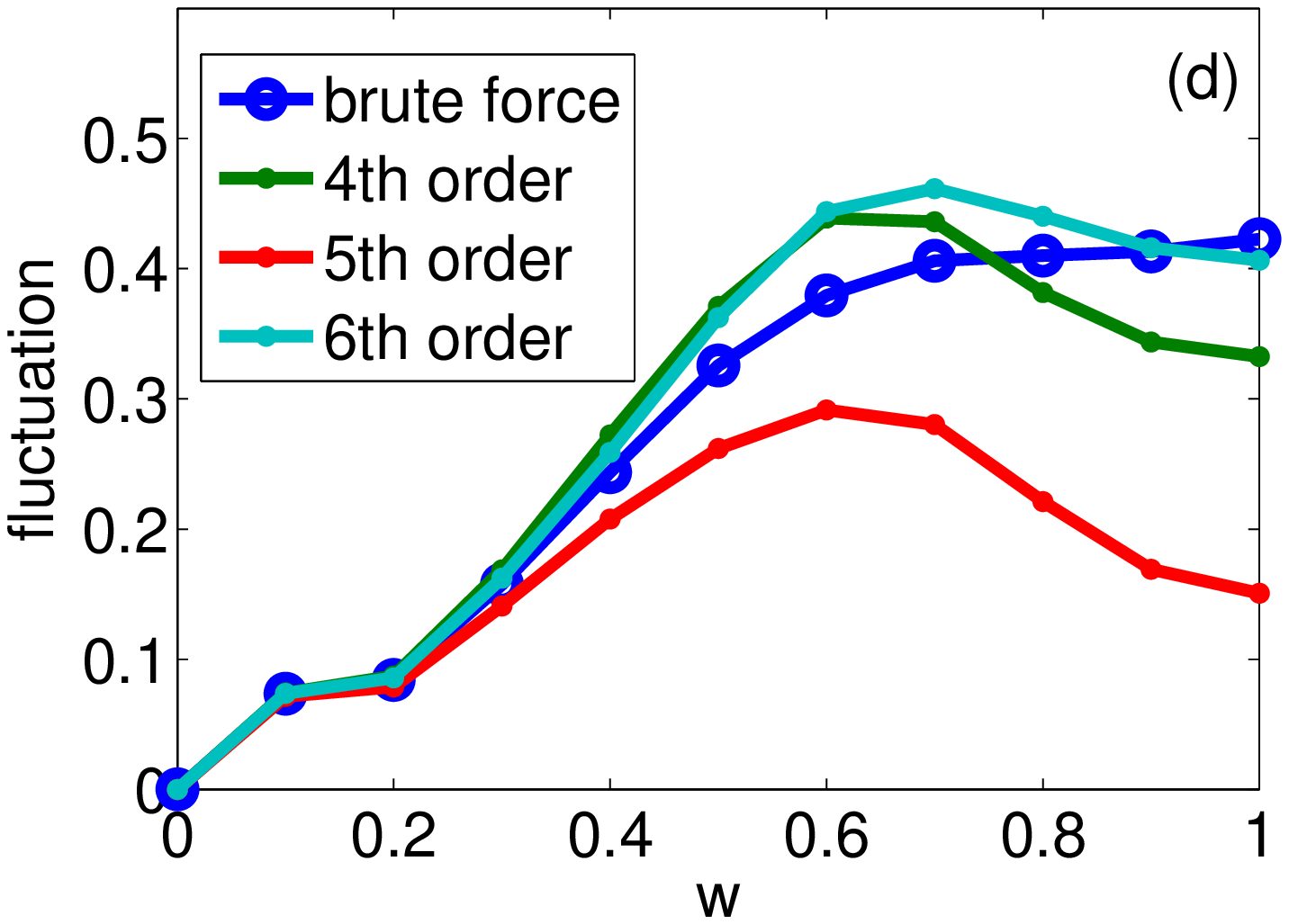}
 \caption{20 layered graphene, 10\% doping , $E=0.55$. (a) Conductance, direct expansion up to different order vs brute force. (b) Conductance, direct expansion up to 6th order vs brute force vs NVC. (c) Averaged shot noise, direct expansion up to different order vs brute force. (d)  Conductance fluctuation, direct expansion up to different order vs brute force.}\label{fig11}%
\end{figure}

Fig.\ref{fig1} - Fig.\ref{fig11} depict our results. Each figure has four panels. In panel (a), we compare our result of average conductance expanded at different orders with that of the brute-force method (blue circle). In the panel (b), we compare our result up to the 6th order with results obtained from the brute-force method as well as the NVC method. The panel (c) and (d) show the averaged shot noise and the conductance fluctuation, respectively, where we compare our results with that of brute-force method. In the brute-force calculation, we have collected $30000$ random configurations for each data point on the curve.

In Fig.\ref{fig1} we show the results on $40 \times 40$ square lattice with Anderson disorder. We have
fixed the Fermi energy to $E=2$ where we have twenty incoming channels. We see from Fig.\ref{fig1}(a) that up
to the 4th or 5th order, our expansion result agrees with that of the brute-force method for disorder strength up
to $W=0.6$. For the 6th order, the good agreement is extended to $W=0.8$. We note that up to $W=1$ the method
of NVC and brute-force give the same result (Fig.\ref{fig1}(b)). For the shot noise (Fig.\ref{fig1}(c)), the
4th and 5th orders seems to give almost the same result and up to $W=0.4$ good agreement is reached. For the
6th order expansion the agreement is better for $W$ up to 0.5. We see that the direct expansion method
underestimate the conductance and overestimate the shot noise. For the conductance fluctuation, the situation
is different. From Fig.\ref{fig1}(d) we see that the conductance fluctuation is of order $2e^2/h$ which is
a well known result in mesoscopic physics. It is interesting to see that the 4th order expansion is better
than 5th and 6th orders. The range of $W$ to have good agreement is $W=0.6$.
One thing to note. Although here we benchmark our result on the lattice model with Anderson disorder, the previous knowledge such as Anderson localization, the universal conductance fluctuation and the percolation theory can not be expected from our approach because those physics require that the strength of disorder large enough and the system enters diffusive and even localization region, but our method can not reach that region due to its perturbative nature.

Now we dope the system with a fixed impurity strength $W$ and two different doping concentrations. For $1\%$ doping (Fig.\ref{fig2}), very good agreement can be obtained for conductance among three methods: NVC, brute-force, and direct expansion up to 6th order in the window of $W=(0,1)$. For the shot noise and conductance fluctuation, 6th order expansion can give good agreement for $W$ up to 1. When we increase the doping concentration, our results deviate from that of the brute-force. At $10\%$ doping concentration (Fig.\ref{fig3}) we find that for average conductance, the range of $W$ decreases to $W=0.7$ while for shot noise and conductance fluctuation the agreement is not good beyond $W=0.3$.

One word on the computational time. In our proposed expansion method, the time cost is dominated by solving the CPA self-consistent equation. As an example, for 2D 40 by 40 lattice model, 10\% doping case, we need 11 steps to obtain the CPA solvent and each step 2.5 seconds. After that, we spend approximately 40 seconds to obtain the fluctuation. However, this time used together can only be used to calculate approximately 50 configurations, from which even the mean value can not be surely given. As the system goes larger, our time advantages becomes more obvious.

We have also studies the average conductance, shot noise, and conductance fluctuation in a disordered graphene ribbon system of size $30 \times 20$ with hard-wall boundary condition perpendicular to the transport direction. Here we use the simplest non-spin tight-binding Hamiltonian on the honeycomb lattice, which is
\begin{eqnarray}
H = \sum_i E_0 a_i^\dagger a_i - \sum_{<ij>}t a_i^\dagger a_j
\end{eqnarray}
In graphene, the nearest hopping energy is $t=2.75\mathrm{eV}$, and we set $t=1$ as the energy unit, then both the Fermi energy and and the disorder strength are measured according to it. Besides, in the above Hamiltonian, $\langle ij\rangle$ denote the nearest neighbor hopping, with the nearest-neighbor unit vector $\mathbf{a}_1=a(0,1)$, $\mathbf{a}_2=a(-\sqrt 3/2,-1/2)$, $\mathbf{a}_3=a(\sqrt 3/2,-1/2)$, and the lattice constant $a=0.142$nm. In the following calculation, we fix the Fermi energy $E_0=0.55$ where there are 15 incoming channels. For Anderson disorder, we see from Fig.\ref{fig9} that for average conductance good agreement is obtained for disorder strength up to $W=0.4$. For shot noise, however, the deviation can be seen when $W=0.3$. To our surprise, the conductance fluctuation from direct expansion method is good for $W$ as large as 0.4. For low doping concentration at $1\%$, Fig.\ref{fig10} shows that good agreement between our method and brute-force method can be reached for average conductance and shot noise with disorder strength up to $W=1$ while for conductance fluctuation reasonable agreement is obtained for $W$ up to $0.8$. For larger doping concentration, the agreement is good for smaller disorder strength. For instance at $10\%$ doping (Fig.\ref{fig11}), the average conductance is good up to $W=0.5$ while for shot noise and conductance fluctuation $W$ is about 0.4 for a reasonable agreement compared with brute-force method.

\section{conclusions}\label{conclusion}
In this paper, we have developed a direct expansion approach to deal with the average shot noise and the conductance fluctuation for disordered systems. Two kinds of disorder were considered: Anderson disorder and the random dopant. We have bench marked our results on a graphene system and a two dimensional square lattice model.
Our results can be summarized as follows.  We find that our expansion method up to the 6th order is comparable, although not as good as NVC method for the calculation of averaged conductance. Up to the sixth order, our results of shot noise and conductance fluctuation agree well with the brute-force method for Anderson impurities with disorder strength up to $W \sim 0.5$ for the square lattice and $W \sim 0.3$ for the graphene system. In the presence of dopant at small doping concentration (1\%) our results are good when $W$ is around 0.9. In general, up to the same order of expansion, average conductance gives better result than the shot noise and conductance fluctuation while the shot noise is the least accurate quantity. One can improve the accuracy by going to higher order expansion at the expenses of more CPU time. Since our method is an expansion approach, it can not deal with large disorder strength and high doping concentration.
Our formalism can be combined with LMTO type of first principles calculation, which can give quantitative prediction to the conductance fluctuation for nano-devices. In the realistic device calculations, such comparisons with brute force method are also in principle available. For example, in the realistic doping devices, one can generate a large number of random configurations at the given concentration, and the averaged shot noise as well as conductance fluctuation can be exactly evaluated. Thus our method is controllable and should be successful as long as CPA itself is valid.

\section*{ACKNOWLEDGEMENT}
The authors would like to thank L. Zhang, G. B. Liu, Y. Wang, Y. Zhu, and H. Guo for their helpful discussions.
We gratefully acknowledge the support from Research Grant Council (HKU 705611P) and University Grant Council (Contract No. AoE/P-04/08) of the Government of HKSAR. This research is conducted using the HKU Computer Centre research computing facilities that are supported in part by the Hong Kong UGC Special Equipment Grant (SEG HKU09).

$^*$ Electronic address: jianwang@hku.hk
%

\end{document}